\documentclass[10pt,a4paper]{article}
\usepackage[left=2cm,right=2cm,top=2cm,bottom=2cm]{geometry}
\usepackage{url}

\usepackage{amsmath,amssymb}
\usepackage{float}

\usepackage{changepage}

\usepackage[utf8]{inputenc}

\usepackage{textcomp,marvosym}

\usepackage{cite}

\usepackage{nameref,hyperref}

\usepackage{microtype}
\DisableLigatures[f]{encoding = *, family = * }

\usepackage[table]{xcolor}
\usepackage{array}

\usepackage{graphicx}
\usepackage{subfigure}
\usepackage{ragged2e}
\usepackage{color}
\usepackage{mathtools}

\usepackage[normalem]{ulem}

\newcolumntype{+}{!{\vrule width 2pt}}

\usepackage[aboveskip=1pt,labelfont=bf,labelsep=period,justification=raggedright,singlelinecheck=off]{caption}

\makeatletter
\renewcommand{\@biblabel}[1]{\quad#1.}
\makeatother

\date{}

\usepackage{lastpage,fancyhdr,graphicx}
\usepackage{epstopdf}

\begin{document}
\begin{center}
{\Huge
\textbf{Fractal and multifractal properties of electrographic recordings of human brain activity} % Please use "sentence case" for title and headings (capitalize only the first word in a title (or heading), the first word in a subtitle (or subheading), and any proper nouns).
\\
\vspace{0.2in}
}

Lucas G. S. Fran\c{c}a\textsuperscript{1\textcurrency*},
Jos\'e G. V. Miranda\textsuperscript{4},
Marco Leite\textsuperscript{1},
Niraj K. Sharma\textsuperscript{1},
Matthew C. Walker\textsuperscript{1},
Louis Lemieux\textsuperscript{1},
Yujiang Wang\textsuperscript{1,2,3\textcurrency*}
\\
\bigskip
\textbf{1} Department of Clinical and Experimental Epilepsy, UCL Institute of Neurology, University College London, London, United Kingdom
\\
\textbf{2} Interdisciplinary Computing and Complex BioSystems (ICOS) research group, School of Computing, Newcastle University, Newcastle upon Tyne, United Kingdom
\\
\textbf{3} Institute of Neuroscience, Newcastle University, Newcastle upon Tyne, United Kingdom
\\
\textbf{4} Institute of Physics, Federal University of Bahia, Salvador, Bahia, Brazil
\\
\bigskip

% Additional Equal Contribution Note
% Also use this double-dagger symbol for special authorship notes, such as senior authorship.

% Use the asterisk to denote corresponding authorship and provide email address in note below.
* lucas.franca.14@ucl.ac.uk and yujiang.wang@newcastle.ac.uk

\end{center}
\section*{Abstract}
The quantification of brain dynamics is essential to its understanding. However, the brain is a system operating on multiple time scales, and characterisation of dynamics across time scales remains a challenge. One framework to study such dynamics is that of fractal geometry. However, currently there exists no established method for the study of brain dynamics using fractal geometry, due to the many challenges in the conceptual and technical understanding of the methods. We aim to highlight some of the practical challenges of applying fractal geometry to brain dynamics, and propose solutions to enable its wider use in neuroscience.

Using intracranially recorded electroencephalogram (EEG) and simulated data, we compared monofractal and multifractal methods with regards to their sensitivity to signal variance. We found that both monofractal and multifractal properties correlate closely with signal variance, thus not offering new information about the signal. However, after applying an epoch-wise standardisation procedure to the signal, we found that multifractal measures could offer non-redundant information compared to signal variance, power (in different frequency bands) and other established EEG signal measures. We also compared different multifractal estimation methods to each other in terms of reliability, and we found that the Chhabra-Jensen algorithm performed best. Finally, we investigated the impact of sampling frequency and epoch length on the estimation of multifractal properties. Using epileptic seizures as an example event in the EEG, we show that there may be an optimal time scale (i.e. combination of sampling frequency and epoch length) for detecting temporal changes in multifractal properties around seizures.

The practical issues we highlighted and our suggested solutions should help in developing a robust method for the application of fractal geometry in EEG signals. Our analyses and observations also aid the theoretical understanding of the multifractal properties of the brain and might provide grounds for new discoveries in the study of brain signals. These could be crucial for understanding of neurological function and for the developments of new treatments.

% Please keep the Author Summary between 150 and 200 words
% Use first person. PLOS ONE authors please skip this step. 
% Author Summary not valid for PLOS ONE submissions.   
\section*{Author summary}
Investigating brain dynamics over several different time scales remains a fundamental challenge of neuroscience. One theoretical framework that has been proposed to tackle this challenge is that of fractal geometry, where signal properties across different time scales are characterised by a single (monofractal) or a range (multifractal) of so-called exponent(s). In this work, we highlight some of the practical pitfalls associated with the application of the fractal geometry framework to electroencephalography (EEG) data. Particularly, we show that signals variance directly impacts the fractal measures, and appropriate pre-processing is required to avoid simply capturing signal variance. We also revealed that certain physiological processes (e.g. epileptic seizures) in the EEG may require a specific timescale to enable their characterisation. Our results have wide-ranging implication for the application of fractal analyses to EEG data.

% Use "Eq" instead of "Equation" for equation citations.
\section*{Introduction}

Brain dynamics are non-linear and are often considered as one of the most complex natural phenomena, involving several different and interacting temporal scales. For example, fast electric activity, slower chemical reactions, and even slower diffusive processes have been observed in the brain. Interestingly, brain dynamics have also been characterised as “scale-free”  \cite{Stam2004,Fraiman2012}, meaning that certain signal properties stay preserved across different time scales. To describe and quantify such (time) scale invariant dynamics, the framework of fractal geometry is often applied.  \\

Fractals have two specific properties: they consist of parts that are similar to the whole -- termed auto-similarity, and they have a fractional Hausdorff-Besicovitch dimension, also called fractal dimension (FD)  \cite{Feder1988,Mandelbrot1983,Falconer2003}. Fractal geometry has been applied to the study of human brain dynamics in health  \cite{Pereda1998,Linkenkaer-Hansen2001,Papo2017,Gong2003} and disease \cite{Gomez2009,Zappasodi2014,Esteller1999}, providing intriguing results. E.g., FD has been shown to vary prior to and during epileptic seizures  \cite{Esteller1999}. Objects adequately characterised by a single fractal dimension are referred to as monofractals. 
However, the fractal formalism has to be extended to capture certain phenomena that cannot be described by a single fractal dimension; these are called multifractals  \cite{Stanley1999}. Multifractal objects can be conceived as decomposable into different subsets or parts, each characterised by its own distinct fractal dimension. Multifractal objects are often described by a spectrum, showing the subsets and their corresponding fractal dimensions. Some natural phenomena exhibit multifractal patterns, for example, turbulence  \cite{Meneveau1987,Chhabra1989}, soil composition  \cite{Vidal-Vazquez2013,Paz-Ferreiro2010b,Paz-Ferreiro2010a,Miranda2006,Zeleke2006,Vazquez2008}, heart beat patterns  \cite{Ivanov1999}, and human physical activity  \cite{Franca2017}. \\

There is also considerable evidence that brain dynamics are multifractal  \cite{Ciuciu2012,Ihlen2010,Zhang2015,Suckling2008,Papo2017,Zorick2013}. At the very least, a breakdown of the monofractal power-law pattern in brain dynamics indicates that additional statistical moments, not reflected in a monofractal characterisation, may be needed to characterise such data  \cite{Fraiman2012}. Furthermore, it is known that interacting processes with different fundamental time scales, similar to those observed in the brain, can generate multifractal patterns  \cite{Argoul1989,Suckling2008}. \\
 
To measure the multifractal spectrum in brain dynamics, Multifractal Detrended Fluctuation Analysis (MF-DFA)  \cite{Kantelhardt2002} is currently the most used approach  \cite{Ihlen2012,Zhang2015}. However, more advanced and potentially more stable estimation techniques have been proposed, such as the Multifractal Detrended Moving Average \cite{Xu2017}, and Chhabra-Jensen approaches \cite{Chhabra1989}. These techniques, to our knowledge, however, have not yet been evaluated with brain signals. In addition, there are several parameter choices to be made for the purpose of the analysis. For example, to capture time-varying changes in multifractal properties, the epoch length and sampling frequency have to be chosen. These parameters may impact the multifractal estimation, but, to date, have not been studied systematically in the context of brain dynamics. \\

The biggest gap in the literature so far, however, is how multifractal properties relate to existing time series signal measures of brain dynamics (e.g. variance of the signal, power spectrum, etc.). A major concern is that complex methods of analysis may not offer a significant advance over simpler, already established methods. For example, in the analysis of the electroencephalogram of epileptic seizures, complex methods were found to actually reproduce patterns detected by simpler measures such as variance of the signal  \cite{Martinerie2003,McSharry2003}. It is therefore essential to understand how the (mono- and multi-) fractal measures relate to more traditional measures, and if new information can be obtained from the signal by applying a mono- or multi- fractal formalism. \\

To summarise, there is a knowledge gap in three critical areas: 1) which (multi)fractal characterisation methodology is best suited for brain signals? 2) what are the optimal estimation parameters (e.g. in terms of recording epoch length) of potentially time varying multifractal properties? 3) what is the relationship between (multi)fractal properties and more traditional and established time series signal measures? 
To address these questions, we chose to analyse intracranially recorded human electroencephalography (icEEG) data, due to its high temporal resolution and high signal to noise ratio. %Initially, we explored the monofractal properties of simulated and icEEG signals, which displayed a strong dependence on signal variance. Thus we turned our attention to multifractal properties of the EEG. We evaluated the reliability of different multifractal estimators, namely, MF-DFA  \cite{Kantelhardt2002}, MF-DMA  \cite{Gu2010} and Chhabra-Jensen \cite{Chhabra1989} on computationally generated time series with known temporally stable multifractal properties. This allowed us to outline an optimal method of signal pre-processing and multifractal spectra estimation, which can be applied to icEEG data. Finally, we have related multifractal properties in icEEG with established EEG time series signal measures, and we also show that there may be a characteristic time scale for detecting multifractal changes around epileptic seizures. \\

\section*{Materials and methods}

%In this section, we first outline four experiments designed to test the performance of estimation methods, as well as to elucidate the impact of multifractal estimation parameters. Then we will describe the two monofractal and three multifractal estimation methods in detail. We then show the generation of time series data with known mono- and multifractal properties, to test the performance of the estimation methods. To test the multifractal measures on real-life brain signals, we then applied our analysis on human intracranial and scalp EEG. Thus, finally, we will summarise the EEG datasets used in this work. \\

To address the questions above, we will first outline four experiments. We will then provide details on monofractal and and multifractal estimation methods, and also show how time series data with known mono- and multifractal properties can be generated to test the performance of the estimation methods. To test the multifractal measures on real-life brain signals, we then applied our analysis on human intracranial and scalp EEG. Thus, finally, we will summarise the EEG datasets used in this work.

The original scripts used in this work are available in \url{https://github.com/yujiangwang/MultiFractalEEG} (to be made publically available upon publication of the manuscript). In addition, the following software packages were used: MATLAB; R  \cite{R2017}; and ggplot2, R.matlab, reshape2, PerformanceAnalytics, and RColorBrewer \cite{ggplot2,Rmatlab,reshape,PerformanceAnalytics,RColorBrewer}. 

\subsection*{Experiments}

\subsubsection*{Experiment 1: Monofractal estimation with respect to changing signal variance}

Estimation of monofractal properties has been applied to EEG signals in the past with varying and often contrasting results \cite{Esteller1999,Li2005}. A particular concern is that complex measures may simply reflect simple properties of the signal \cite{Martinerie2003,McSharry2003}.  Hence, in our first analysis, we focus on the relationship between monofractal measures and signal variance. For this, we used a simulated monofractal time series (termed fractional Brownian motion, or short fBm) with its standard deviation modulated by a modified ramp function. 

The fBm was simulated with a Hurst exponent $H = 0.7$ and a modulating function M (described in more detail later and in Suppl. S1 App) and split into 1800 1024-sample epochs.  We estimated the monofractal dimension of this simulated signal using the Higuchi and DFA methods. To assess the impact of signal variance, we have also tested the effect of epoch-based standardisation. To ensure that our effects were not simply an artifact generated by the fBm, we also repeated the analysis on one exemplary icEEG data segment.

\subsubsection*{Experiment 2: Multifractal estimation stability}

As the monofractal analysis proved unsatisfactory, we then turned our attention to multifractal analysis. 
In order to evaluate the stability of multifractal properties in time, we generated a time series exhibiting stable multifractal properties over time using the p-Model. The time series was then evaluated using an epoch-based approach with the three estimators: MF-DFA, MF-DMA, and Chhabra-Jensen. The stability of the estimator can then simply be assessed as the temporal variability of its output.

\subsubsection*{Experiment 3: Multifractal estimation of human EEG and its potential added value}

To assess whether the chosen multifractal metrics contribute any information about the signal in addition to more established signal metrics, we analysed human EEG signals recorded intracranially. Again, we used an epoch-based approach, and we compared the multifractal metrics to a number of conventional signal metrics (mean, standard deviation, line length, bandpower) on each epoch. The similarity between metrics was evaluated using Pearson correlation and Mutual Information (the code is available at \url{https://github.com/robince/gcmi}) \cite{Ince2017}. Furthermore, monofractal metrics were also included in this comparison, to further demonstrate the advantages in applying a multifractal over monofractal approaches.

\subsubsection*{Experiment 4:  Impact of sampling frequency and epoch length on multifractal estimation of human EEG}

Finally, we also evaluated the impact of the multifractal estimation parameters in the characterisation of a seizure. We used intracranial EEG signals recorded from four patients undergoing pre-surgical planning, the signals were originally sampled at 5000 Hz. For this analysis, down-sampled versions were evaluated with epochs of different sizes. To assess the effect of sampling frequency, we down-sampled the signal to 4000 Hz, 3000 Hz, 2500 Hz, 2000 Hz, 1000 Hz, 800 Hz, 750 Hz, 600 Hz, 500 Hz, 400 Hz, 300 Hz, and 250Hz. For each sampling frequency, we evaluated different epoch sizes (1024 points, 2048 points, 4096 points, 8192 points, and 16384 points).

We defined a difference in multifractal spectrum width ($\Delta\alpha^{\dagger}$) during the seizure compared to the background as the effect size (Cohen’s D) between the ictal and interictal periods:

\begin{equation}
D = \dfrac{<\Delta\alpha^{\dagger}_{ictal}> - <\Delta\alpha^{\dagger}_{interictal}>}{s(\Delta\alpha^{\dagger}_{interictal})}
\end{equation}

where $<\Delta\alpha^{\dagger}>$ represents the mean and s denotes standard deviation.

\subsection*{Fractal dimension estimation}
To estimate the monofractal dimension from a time series, we used two established estimation approaches: Higuchi method \cite{Higuchi1988} and Detrended Fluctuation Analysis \cite{Peng1994}. These methods are widely applied in the literature and aim to capture the features of a time series in a single scaling exponent.

\subsubsection*{Higuchi method}

The Higuchi method consists of constructing series with elements of an original time series and measuring their lengths \cite{Higuchi1988}. Given a time series with $N$ time points $X(1),X(2),...,X(N)$, the equation \ref{eq:resampleHiguchi} shows a rule for reconstructing smaller time series with elements of the original recording. The lengths of the time series can be assessed according to equation \ref{eq:higuchi}. The brackets $\lfloor\rfloor$ represent Gauss’ notation, i.e., the rounded integer of the division  \cite{Higuchi1988}. The variable $d$ represents a down-sampling factor of the original time series.

\begin{equation}
X(m),X(m+d),X(m+2d),...,X\bigg(m+\bigg\lfloor\dfrac{N-m}{d}\bigg\rfloor d\bigg) \quad \text{where } \quad m=1,2,...,d
\label{eq:resampleHiguchi}
\end{equation}

\begin{equation}
L_{m}(d)=\dfrac{\Bigg\{\sum_{i=1}^{[(N-m)/d]}|X(m+id)-X(m+(i-1)d)|\dfrac{N-1}{\lfloor(N-m)/d\rfloor d}\Bigg\}}{d}
\label{eq:higuchi}
\end{equation}

If the average curve length $<L_{m}(d)>_{m}$ over $d$ sets follows a power law, according to equation \ref{eq:higuchiFD}, the time series has scaling properties, with a fractal dimension $FD_{Hig}$.  \\

\begin{equation}
<L(d)> \propto d^{-FD_{Hig}}
\label{eq:higuchiFD}
\end{equation}

The routine used in the estimation of Higuchi fractal dimension FD is available at \url{https://uk.mathworks.com/matlabcentral/fileexchange/50290-higuchi-and-katz-fractal-dimension-measures}.

\subsubsection*{Detrendred Fluctuation Analysis} 

The Detrended Fluctuation Analysis (DFA) method is an alternative method  \cite{Peng1994,Peng1995}, which estimates the Hurst exponent $H$ in time series data instead of the fractal dimension. In general, the fractal dimension $FD$ will represent local features of the signal whereas the Hurst exponent will reflect on global properties of the time series \cite{Gneiting2001}. However, if the time series is self-similar, the relationship of $FD + H = 2$ can be established. Hence, we will use the DFA as another method of estimating the fractal dimension here. \\

The method consists of the following steps: Initially the time series with $N$ time points $X(1),X(2),...,X(N)$ is integrated as follows:

\begin{equation}
y(k) = \sum_{i=1}^{k} (X(i) - \langle X \rangle)
\label{eq:integratedfa}
\end{equation}

Where $X(i)$ represents the $i-th$ element of the time series and $<X>$ denotes the mean over the whole recording. The second step consists of dividing the time series into $N_l$ windows of length $l$, then the mean square root of the integrated series is subtracted from the local trend, in every window \cite{Peng1995}, as shown in equation \ref{eq:dfa}.

\begin{equation}
F(l) = \sqrt{\frac{1}{N_l} \sum_{k=1}^{N_l} [y(k) - y_l(k)]^{2}}
\label{eq:dfa}
\end{equation}

The local trend ($y_l(k)$) is obtained from a linear regression over the time series in the window, and number $N_l$ represents the total number of windows. In the following step, equation \ref{eq:dfa} is obtained for several window lengths ($l$). The relation between $F(l)$ and $l$ is described by a power law, according to equation \ref{eq:hurst}, where $H$ is the Hurst exponent.

\begin{equation}
F(l) \propto l^{H}
\label{eq:hurst}
\end{equation}
 
The code used here is available in the Physionet repository (\url{https://www.physionet.org/physiotools/dfa/}) \cite{Goldberger2000,Peng1995}.

\subsection*{Multifractal spectrum estimation}

In this section, we describe three multifractal spectrum estimators: Multifractal Detrended Moving Average  \cite{Gu2010}, Multifractal Detrended Fluctuation Analysis  \cite{Kantelhardt2002,Ihlen2010,Ihlen2012}, and Chhabra-Jensen \cite{Chhabra1989}, as these are the most established methods used in the literature. \\

\begin{figure}[H]
\includegraphics[width=\textwidth]{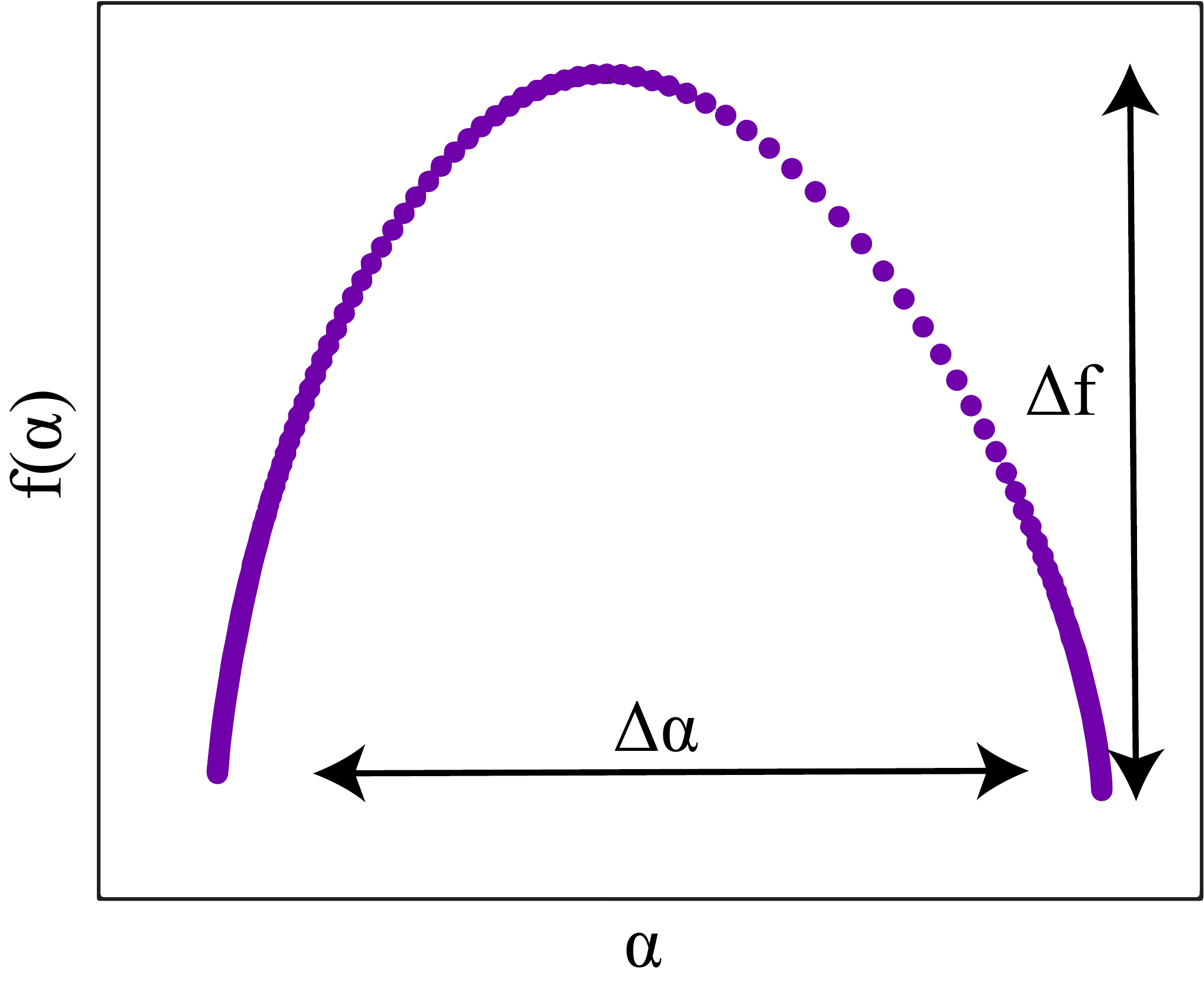}
\caption{{\bf Multifractal singularity spectrum, $f(\alpha)$, with a characteristic parabolic shape.} The spectrum width ($\Delta\alpha$) and height ($\Delta f$) measures are indicated by the arrows.
}

\label{fig:spectrum}
\end{figure}

Multifractal properties are represented as spectra (\ref{fig:spectrum}), where essentially the fractal scaling properties, or more precisely Hausdorff dimensions (often noted as $f(\alpha)$), are measured over a range of different singularities ($\alpha$). Formally, the singularity spectrum is a function that describes the Hausdorff dimension of subsets of the time series $X(t)$ with a specific Hölder exponent, according to: 

\begin{equation}
f(\alpha) = D_{F}\{X(t_s),  H(X(t_s)) = \alpha\}.
\label{eq:sing_spectrum}
\end{equation}

Essentially, $f(\alpha)$ is the Hausdorff dimension ($D_F$) of the subset ($t_s$) of the time series $X(t_s)$ that has a the Hölder exponent $\alpha$ \cite{VandenBerg1999,Murcio2015}.

To characterise the function, or singularity spectrum $f(\alpha)$, usually, the width ($\Delta\alpha$) and height ($\Delta f$) of the spectrum are used to characterise. $\Delta\alpha$ indicates the range of singularities present in a time series, this is also the most commonly used measure of how multifractal an object is. The spectrum height $\Delta f$ indicates the range of Hausdorff dimensions present in the time series. See Fig. \ref{fig:spectrum} for an exemplary singularity spectrum plot.

\subsubsection*{MF-DMA} 

Multifractal Detrended Moving Average (MF-DMA) is one of the most commonly used methods for the estimation of multifractal measures. The method of calculation consists of the following steps \cite{Gu2010}: Given time series $X(t)$ with time points $X(1),X(2),...,X(N)$, the cumulative sum time series is obtained:

\begin{equation}
y(t) = \sum_{t=1}^{N}X(t)
\end{equation}

We then calculate the moving average over time window of length $l$:

\begin{equation}
\tilde{y}(t) = \frac{1}{l}\sum_{z=0}^{l-1}y(t-z)
\end{equation}

A detrended version of the signal is obtained by the subtraction:

\begin{equation}
\epsilon(i) = y(i) - \tilde{y}(i)
\end{equation}

The resulting series is then divided in $N_l$ disjoint sets of points of size $l$ and a root-mean-square function is obtained for each epoch $\nu$ via:

\begin{equation}
F_{\nu}(l) = \bigg\{\frac{1}{l}\sum_{i=1}^{l}\epsilon_{\nu}^{2}(i)\bigg\}^{\frac{1}{2}}
\end{equation}

A generalised $q$th-order overall fluctuation function can be obtained from:

\begin{equation}
F_q(l) = \bigg\{\dfrac{1}{N_{l}}\sum_{\nu=1}^{N_{l}}F_{\nu}(l)^{q}\bigg\}^{\frac{1}{q}} \quad q \neq 0
\end{equation}

and

\begin{equation}
\ln{F_0(l)} = \frac{1}{N_{l}}\sum_{\nu=1}^{N_{l}}\ln F_{\nu}(l) \quad \text{for} \quad q = 0
\end{equation}

It is possible to find a power-law relationship between $F^q(l)$ and the window length, or scale $l$ by:

\begin{equation}
F^q(l) \propto l^{\alpha(q)} 
\label{eq:plaw}
\end{equation}

The multifractal ``mass exponent'' \cite{Biswas2012} can be defined as:

\begin{equation}
\tau(q) = q\alpha(q) - D_{f}
\label{eq:tau}
\end{equation}

where $D_{f}$ is the fractal dimension of the support measure. For a single-channel time series, $D_{f} = 1$. The spectrum, $f(\alpha)$, can be obtained with a Legendre transform \cite{Gu2010}:

\begin{equation}
\alpha(q) = \frac{d\tau(q)}{dq}
\end{equation}

\begin{equation}
f(q) = q\alpha - \tau(q)
\end{equation}

\subsubsection*{MF-DFA} 

The Multifractal Detrended Fluctuation Analysis (MF-DFA) method is essentially a generalisation of the DFA approach \cite{Kantelhardt2002,Ihlen2012}. The time series is first rebuilt according to eq. \ref{eq:integratedfa}.

It is then divided into $N_{l} = \frac{N}{l}$ non-overlapping epochs $\nu$ of length $l$. The variance of the detrended series is calculated as follows:

\begin{equation}
F^2_{\nu}(l) = \frac{1}{l}\sum_{k=1}^{n}(y((\nu - 1)l + 1) - y_{v}(k))^2
\label{eq:detvar}
\end{equation}

where $y_{\nu}$ represents the fitting in the epoch $\nu$ obtained via linear regression. The overall q-th order fluctuation functions can be obtained as:

\begin{equation}
F_{q}(l) = \bigg\{\frac{1}{N_{l}}\sum_{\nu=1}^{N_{l}}(F^2_{\nu}(l))^{\frac{q}{2}}\bigg\}^{\frac{1}{q}}
\label{eq:bgDFA}
\end{equation}

A log-log plot of $F_{q}(l)$ versus $l$ for different values of q should present a linear curve defined by the power law in equation \ref{eq:plaw}. Similarly to the MF-DMA method, the multifractal scaling exponent can be defined as in \ref{eq:tau} and the spectrum $f(\alpha)$ can be determined in the same way as in the MF-DMA approach.

\subsubsection*{Chhabra-Jensen} 

Multifractal spectra can be obtained in a more direct way, without the need for the Legendre transform using the Chhabra-Jensen (CJ) method \cite{Chhabra1989,Franca2017,Murcio2015,Vidal-Vazquez2013,Paz-Ferreiro2010b,Paz-Ferreiro2010a,Miranda2006,Zeleke2006,Vazquez2008,Xu2017}. 
Considering a time series as a distribution over time, the approach consists of calculating a family of generalised measures by covering the time series with windows. These are probabilistic measures with an emphasis factor $q$ that accentuates different singularities depending on its value. More singular regions are emphasised by $q > 1$ whereas less singular regions will have a higher weight with $q < 1$ \cite{Chhabra1989}.

First, we define:
\begin{equation}
\mu_{i}(q,l) = \frac{P_{i}(l)^{q}}{\sum_{j}P_{j}(l)^{q}}
\label{eq:genMeasures}
\end{equation}

where $P_{i}(l)$ represents the cumulative probability of a window $i$. $l$ corresponds to the size of the window in which the generalised measures are obtained. The window epochs are indexed by the variables $i$ and $j$. Then the multifractal spectra can be obtained directly from:

\begin{equation}
    \alpha(q) = \lim_{l\rightarrow0}\frac{\sum_{i}{\mu_{i}(q,l)\log{P_{i}(l)}}}{\log{l}}
    \label{eq:alpha}
\end{equation}

and

\begin{equation}
    f(q) = \lim_{l\rightarrow0}{\frac{\sum_{i}{\mu_{i}(q,l)\log{\mu_{i}(q,l)}}}{\log{l}}}
    \label{eq:f}
\end{equation}

A numerical approximation to the equations above is provided by the measures $M\alpha$ and $Mf$ functions in eq. \ref{eq:malpha} and \ref{eq:mf}.

\begin{equation}
M\alpha = \sum_{i}\mu_i(q,l)\log{P_{i}(l)}
\label{eq:malpha}
\end{equation}

\begin{equation}
M\alpha = \sum_{i}\mu_i(q,l)\log{\mu_i(q,l)}
\label{eq:mf}
\end{equation}

$\alpha$ and $f(q)$ can then be obtained as the slopes by regressing these two measures against the scales $l$: $M\alpha \sim l$ and $Mf \sim l$. 

The algorithmic summary of the Chhabra-Jensen method consists of the following steps:  \\

\begin{itemize}

	\item The algorithm has as input the time series, a range of $q$ values to which the spectrum will be evaluated, and window sizes $l$ that vary in a dyadic scale.

	\item The time series is divided into non-overlapping epochs of length $l$ and the generalised measures are estimated according to equation \ref{eq:genMeasures}.
	
	\item The measures $M\alpha$ and $Mf$ are obtained from the generalised measures.
	
	\item $\alpha$ and $f(q)$  in eq. \ref{eq:alpha} and \ref{eq:f}, respectively, are obtained with a linear regression procedure: $\log(M\alpha)$ is regressed against $-\log(l)$ and $log(Mf)$ is regressed against $-\log(l)$, they give $\alpha$ and $f$ respectively as the slopes. 
	\item A rejection criterion is also used, where all $q$ exponent values with $R^2 < 0.9$ in the linear regression are not considered. 

\end{itemize}

The code used in this study to calculate the multifractal spectrum is available at: \url{https://github.com/lucasfr/chhabra-jensen}. A flow-chart diagram of the algorithm is included in Supp. S1 Fig.

\subsection*{Data}

\subsubsection*{Simulating fractal time series: Modulated fractional Brownian motion}

To fully test methods of estimating the monofractal dimension from time series, we computationally produced time series that are known to be fractal (used for Experiment 1). We generated fractional Brownian motion (fBm) profiles/time series using a novel modified version of the Wood-Chan or circulant embedding approach  \cite{Kroese2015,Shevchenko2014} that allow us to change the variance of the signal over time, in order to evaluate its influence on the fractal estimation. Our modulated fBM approach uses a modulating function, $M(t)$, which produces a signal that has an amplitude varying over time. The details of fBm and our Modulated fBm (ModfBm) are described in Suppl. S1 Appendix. The fBm time series was simulated with Hurst exponent $H = 0.7$; the value was chosen due to its persistent features, i.e., it generates a time series with memory. The modulating function $M(t)$ used to modify the variance of the signal over time (see also described in Suppl. S1 Appendix) is shown in Fig. 2(C). Using this method, we generated time series to evaluate the impact of variance change on monofractal estimators. \\

\subsubsection*{Simulating multifractal time series: p-model}

Similarly to the fBm, we also used a computational procedure to generate time series that are known to be multifractal (for Experiment 2) based on the \textit{p-model}, which was developed to reproduce features observed in turbulence experiments known to have multifractal properties \cite{Meneveau1987}. This is a simple model, having a single fraction $p_{1}$ as its only input. Briefly the algorithm works as follows: From an interval of length $L$ and height $\epsilon_{L} = c \quad \text{(} \text{ is a constant)}$, we create two segments of length $L/2$. Based on the input parameter $p_{1}$, it is possible to establish a second fraction in which a second parameter will be given by $p_{2} = 1-p_{1}$. The heights of each interval will thus be given by $y = 2p_{1}\epsilon_{L}$, and $y = 2p_{2}\epsilon_{L}$, respectively. This procedure is repeated for each remaining segment, selecting left or right for $p_{1}$ randomly \cite{Meneveau1987}. \\

We employed the \textit{p-model} in the simulation of a time series profile with multifractal properties to be evaluated by different estimation methods. It was generated with a code available at \url{http://www2.meteo.uni-bonn.de/staff/venema/themes/surrogates/pmodel/} \cite{Venema2006,Davis1997}. Using this algorithm, we generated time series to evaluate the performance of different multifractal estimators with $p = 0.4$.

\subsubsection*{Human EEG data}

Intracranial EEG data segments extracted from recordings in patients undergoing evaluation for epilepsy surgery were used for experiments 3 and 4. In order to evaluate the effect of EEG signal variance change on multifractal properties (Experiment 3), we specifically looked for one recording, where the signal variance changes dramatically over time. One such recording was found in one patient (male, 28 years old, temporal lobe epilepsy, recorded at the National Hospital for Neurology and Neurosurgery (UCLH NHS Foundation Trust, Queen Square, London, UK), patient ID: ‘NHNN1’) near one seizure event. We used a 60-minute recording segment around the epileptic seizure for our analysis. The seizure onset and offset were marked by expert clinicians, independent of this research project. Note that we used this segment specifically due to the dramatic change in signal variance, which actually occurs before the seizure and evolves over about 15 minutes. We do not make conclusions about the seizure event itself at this stage, but rather use this recording as an example to illustrate a technical point about multifractal property estimation from EEG.  \\

To analyse the possible changes in multifractal properties during seizures (Experiment 4), we used a different dataset: Intracranial EEG from four subjects were retrieved from the ieeg.org repository (\url{http://www.ieeg.org/}) \cite{Wagenaar2013}: ’I001\textunderscore P005\textunderscore D01’, ’I001\textunderscore P034\textunderscore D01’, 'I001\textunderscore P010\textunderscore D01', and ’Study 040’. These subjects were chosen due to the high sampling rate of their recordings (5 kHz), as we evaluated the impact of sampling frequency on multifractal properties. We extracted a 30-minute segment around every seizure in each patient for further analysis. In Experiment 4, we performed the multifractal analysis on channels that were marked as seizure onset channels. We show the results for one patient in the main figure and the results for the remaining three patients are shown in Suppl. S2 Appendix. Further information on the recordings is available in Suppl. S1 Table.

\subsection*{Pre-processing and analysis of time series}

Unless stated otherwise, we have applied the same pre-processing and analysis parameters to the computationally generated time series and the human EEG recordings and performed the fractal and multifractal estimations on 1024-sample epochs. 
In experiments 1 and 2, we were specifically interested in the effect of signal variance on the (multi) fractal estimation, and therefore we compared the results for signal subjected to a standardisation procedure, as follows:

\begin{equation}
x' = \frac{X - <X>}{s}
\label{eq:norm}
\end{equation}

where $<X>$ is the epoch mean and $s$ the epoch standard deviation of the time series $X$, resulting in a time series with zero mean and unit standard deviation. \\

The Chhabra-Jensen method requires as input a distribution function over the domain of positive real numbers, which is incompatible with EEG data which contain positive and negative values. Hence, we propose the use of a sigmoid-transformation here (equation \ref{eq:sigmoid}) to map the time series onto positive values, in order to apply the Chhabra-Jensen method. Example sigmoid functions and correspondingly transformed EEG signal are shown in Suppl. S2 Fig.

\begin{equation}
\sigma(X) = \frac{1}{1 + e^{vX}}
\label{eq:sigmoid}
\end{equation}

The parameter $v$ was chosen based on its effect on the estimated multifractal width for three types of time series: icEEG (NHNN1 - channel 1), surrogate EEG (temporally shuffled values of the original time series from NHNN1 - channel 1) and a simulated random series (with the same mean and variance), across the range $v=[0.1, 2.0]$ in steps of 0.1. To find the optimal value for the parameter $v$, we needed to balance the trade-off between the three series in terms of presenting the most distinct $\Delta \alpha$ values (Suppl. S3 Fig (A)), while showing minimum distortion on the recording, or maximum correlation with the original time series (Suppl. S3 Fig (B)). We chose $v = 1$ as an acceptable trade-off point. \\
Finally, to compare multifractal properties to classical EEG frequency band power, we used the following definitions for the classical EEG frequency bands: $\delta$ (0.5-4 Hz), $\theta$ (4-8 Hz), $\alpha$ (8-15 Hz), $\beta$ (15-30 Hz), and $\gamma$ (30-60 Hz).

\section*{Results}

\subsection*{Experiment 1: Monofractal estimation with respect to changing signal variance}

We evaluated the relationship between monofractal measures and signal variance using a simulated time series based on fractional Brownian motion (fBm), where its signal variance is modulated by a modified ramp function. The modulation function is shown in Fig. 2(A) and resulting time series in Fig. 2(B). The standard deviation of the generated time series indeed tracks the shape of the modulating function (Fig. 2(C)). \\

We estimated the monofractal dimension of this simulated signal using two standard methods: Higuchi and DFA. We observe that both methods appear to be affected by the changing signal variance (Fig. 2(D), (F)). Furthermore, the effect persists after epoch-based standardisation (Fig. 2(E), (G)): the monofractal properties and standard deviation correlate with $\rho  = 1.00$ and $\rho  = 0.99$ for the Higuchi and DFA methods, respectively. A similar effect was observed for a real icEEG recording that contained changes in signal variance over time (Suppl. S4 Fig.).

\begin{figure}[H]
\includegraphics[width=1\textwidth]{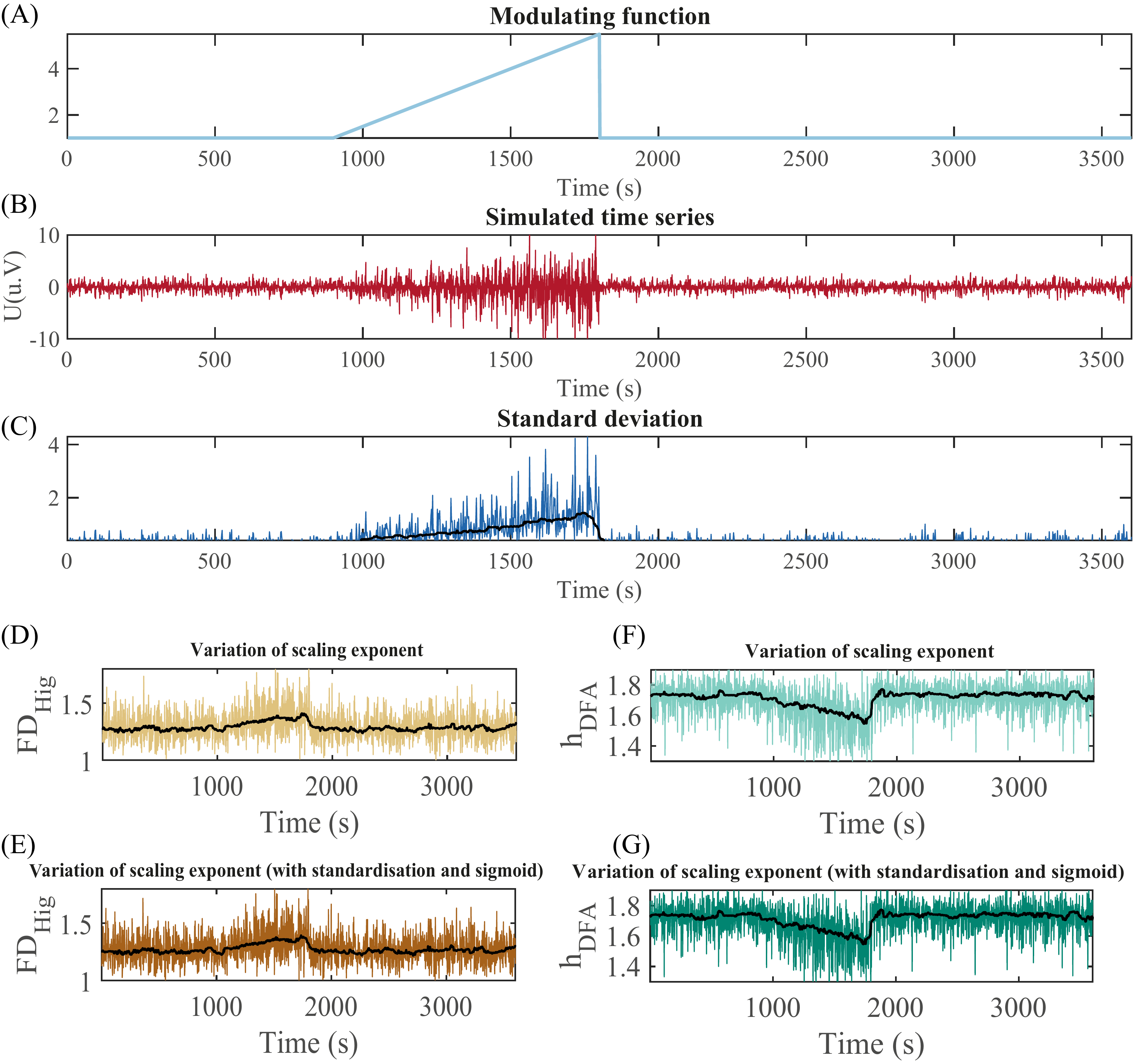}
\caption{{\bf Impact of the signal standard deviation on monofractal scaling exponent estimation.} (A) Modulation of the standard deviation of the time series over time; (B) Time series simulated using fractional Brownian motion based on modulation in A; (C) Standard deviation of the simulated signal in B. (D) Monofractal dimension obtained with the Higuchi method from signal without epoch-based standardisation. (E) Monofractal dimension obtained with the Higuchi method from epoch-based standardised signal.(F) Hurst exponent obtained with the DFA method from signal without epoch-based standardisation.  (G) Hurst exponent obtained with the DFA method from epoch-based standardised signal. 
}
\label{fig:stdImpact}
\end{figure}

\subsection*{Experiment 2: Multifractal estimation stability}

Given that monofractal measures are dominated by the signal variance, and therefore not more informative for human EEG characterisation, we decided to turn our attention to multifractal methods. In the following, we will denote the epoch-wise estimates of multifractal width $\Delta\alpha$ and height $\Delta f$ (and $\Delta\alpha^{\dagger}$ and $\Delta f^{\dagger}$ for the epoch-based standardised measures). \\

This experiment was designed to assess the reliability, or stability of the different multifractal estimation methods over time (note that the accuracy of these methods has been demonstrated elsewhere \cite{Kantelhardt2002,Gu2010,Chhabra1989}). Figure 3 shows the simulated signal by the p-model and the outputs of the three multifractal spectral estimation methods. In all cases, the magnitude of $\Delta\alpha^{\dagger}$ and $\Delta f^{\dagger}$ were clearly different from zero, consistent with multifractal behaviour. The ($\Delta\alpha^{\dagger}$, $\Delta f^{\dagger}$) output variances over time for the MF-DFA, MF-DMA, and Chhabra-Jensen estimation methods were: (0.018, 0.18), (4.17e-4, 0.0028) and (2.3e-30, 6.5e-30), respectively. In addition, the MF-DFA output violated the theoretical topological limit of $\Delta f^{\dagger}$ = 1. As the Chhabra-Jensen method shows the lowest variance over time (i.e. most reliable/stable), it will be our multifractal analysis method of choice for the remainder of this work.

\begin{figure}[H]
\includegraphics[width=\textwidth]{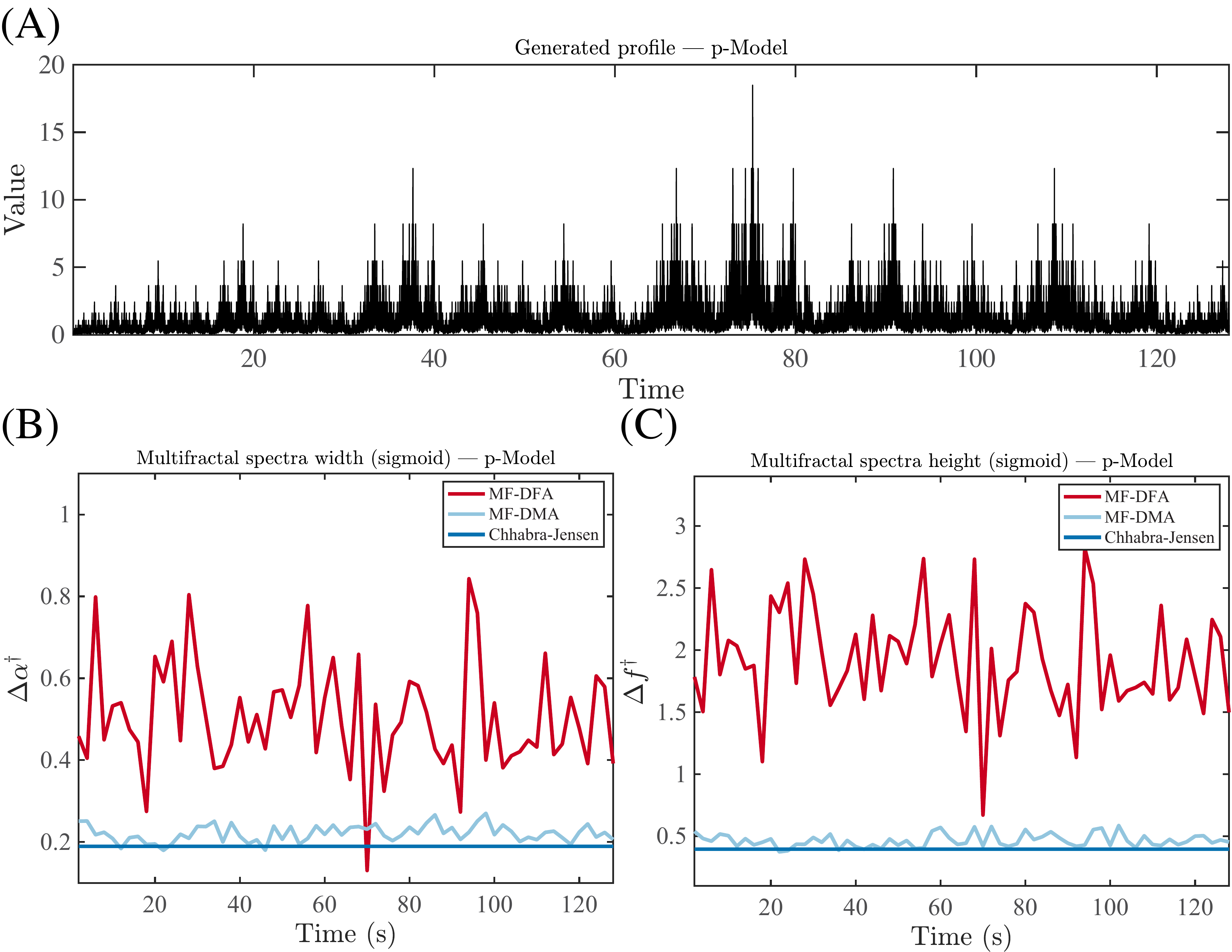}
\caption{{\bf Comparison of three multifractal spectrum estimation methods (MF-DFA, MF-DMA and Chhabra-Jensen) for p-Model simulated time series.} (A) Time series simulated for $p = 0.4$. (B) Estimated multifractal spectra width $\Delta\alpha^{\dagger}$  and (C) height $\Delta f^{\dagger}$.
}
\label{fig:methImpact}
\end{figure}

\subsection*{Experiment 3: Multifractal estimation of human EEG and its potential added value}

Next, we evaluated the relationship between multifractal signal properties and other widely used conventional EEG measures (such as signal variance). Figure 4 shows the results of the multifractal spectrum and conventional measures in comparison. The pattern of multifractal spectrum width without epoch-based standardisation ($\Delta\alpha$) reflects the signal variance closely, in contrast to the estimate for the epoch-based standardised signal ($\Delta\alpha^{\dagger}$). Finally, signal line length also shows a very different temporal profile from $\Delta\alpha^{\dagger}$. A similar figure showing the variation of $\Delta f$ and $\Delta f^{\dagger}$ metrics is available in Suppl. S5 Fig.

\begin{figure}[H]
\includegraphics[width=\textwidth]{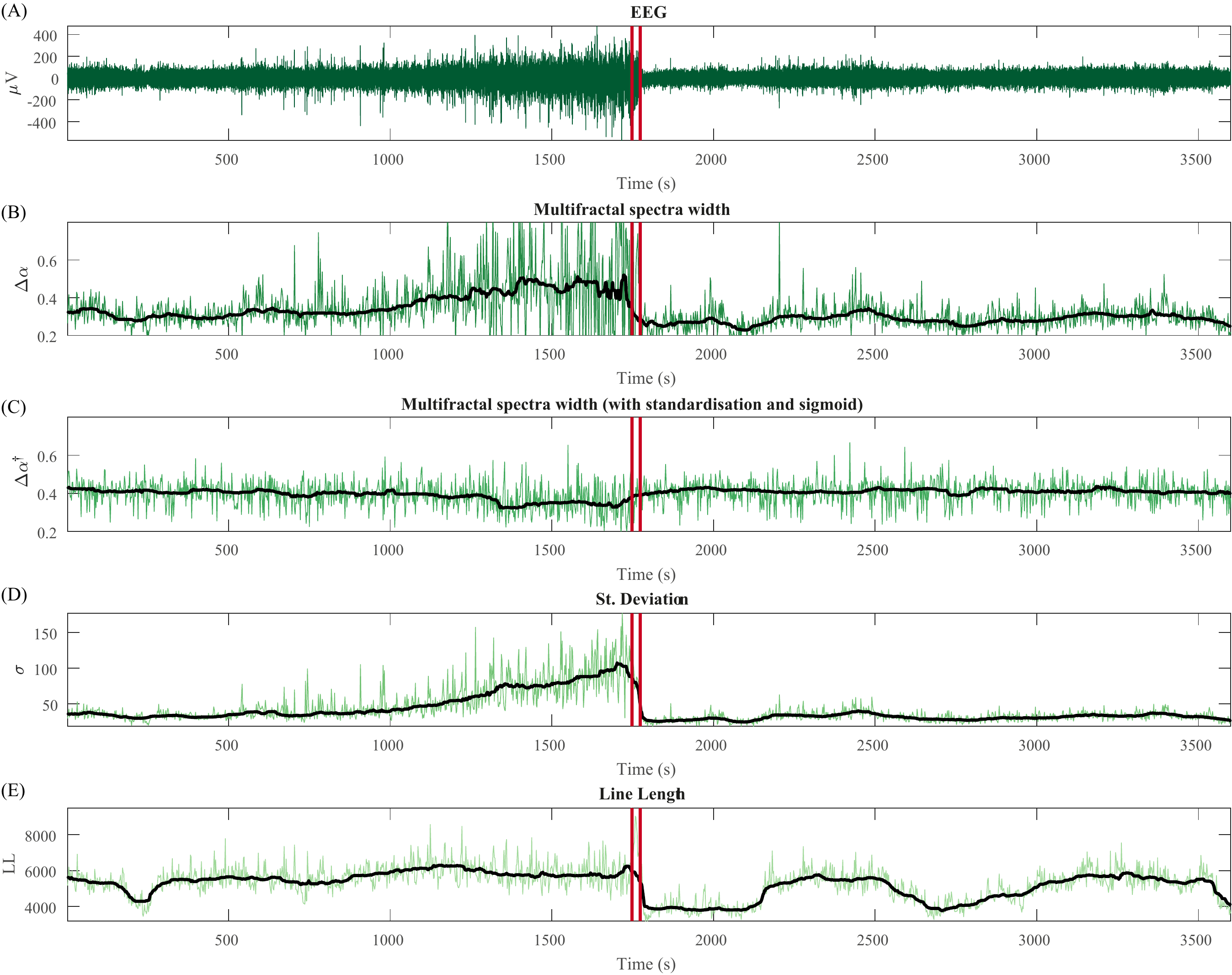}
\caption{{\bf Temporal dynamics of multifractal spectrum width compared with conventional measures for human intracranial EEG.} (A) Intracranial EEG segment containing a seizure (onset and offset marked with red vertical lines). Note that this recording was chosen because it showed a dramatic change in signal variance during non-seizure periods, not because of any seizure related properties. (B): Variation of multifractal spectrum width without epoch-wise standardisation ($\Delta\alpha$). (C): multifractal spectrum width based on epoch-wise standardised time series ($\Delta\alpha^{\dagger}$). (D): Standard deviation (E): Line length. Black line: moving average of each measure.}
\label{fig:szStd}
\end{figure}

Figure 5 shows the quantification of similarities of the signals in Fig. 4 through a correlation analysis. In summary, a high degree of correlation is present between the signal standard deviation, multifractal spectrum width ($\Delta\alpha$), and detrended fluctuation analysis (monofractal approach) both with and without epoch-based standardisation. We found that standardisation reduces the correlation between $\Delta\alpha$ and the standard variation from $\rho = 0.86$ (for $\Delta\alpha$ ) to $\rho = -0.14$ (for $\Delta\alpha^{\dagger}$). We also note that $\Delta\alpha$ is highly correlated with DFA and DFA$^{\dagger}$ estimates ($\rho = 0.74$ and $\rho = 0.71$, respectively) while it is markedly reduced for $\Delta\alpha^{\dagger}$ ($|\rho| < 0.3$). The analysis based on the mutual information \cite{Ince2017} rather than correlation showed a similar pattern (Suppl. S6 Fig.).

\begin{figure}[H]
\includegraphics[width=\textwidth]{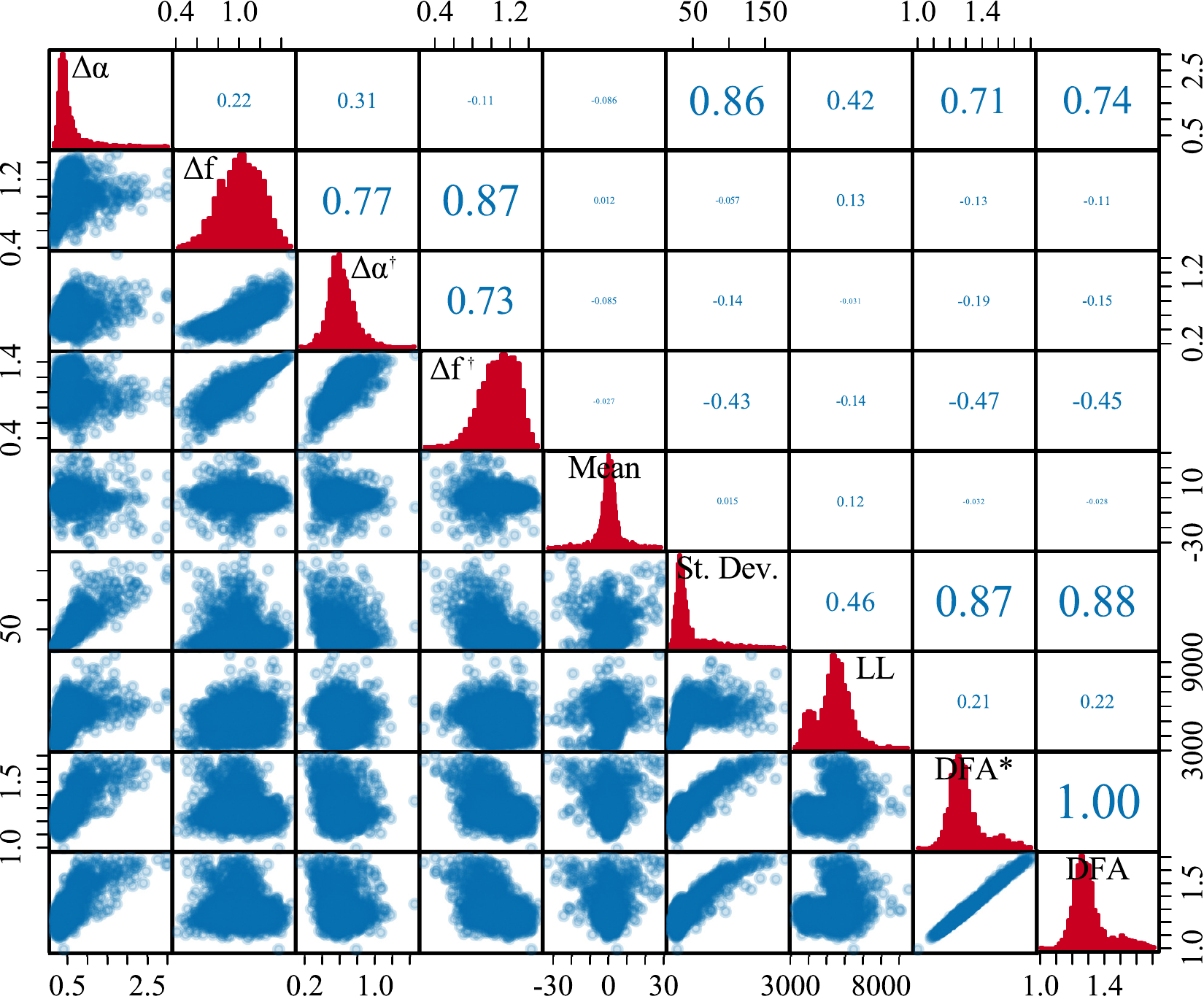}
\caption{{\bf Correlation between multifractal spectrum and conventional EEG measures for human data.} The diagonal of the matrix shows the distribution for each measure across epochs. The lower triangle contains the scatter plots for each pair of measures across epochs. The upper triangle shows the Pearson correlation value for each pair of measure, where the size of the font additionally corresponds to the correlation coefficient to provide an additional visual cue. }
\label{fig:otherMeasures}
\end{figure}

The relationships of the multifractal properties and specific EEG frequency band power are shown in Fig 6. In summary, the correlation values between the multifractal measures $\Delta\alpha^{\dagger}$, $\Delta f^{\dagger}$ and signal power in the classical EEG bands are low ($|\rho| < 0.3$). A supplementary analysis of EEG time series data containing different sleep stages (which are known to be dominated by specific frequencies) shows similar results (see Supplementary material S3 Appendix). Based on these results, we focused on $\Delta\alpha^{\dagger}$ (using epoch-wise standardisation of the time series) in the subsequent analyses.

\begin{figure}[H]
\includegraphics[width=\textwidth]{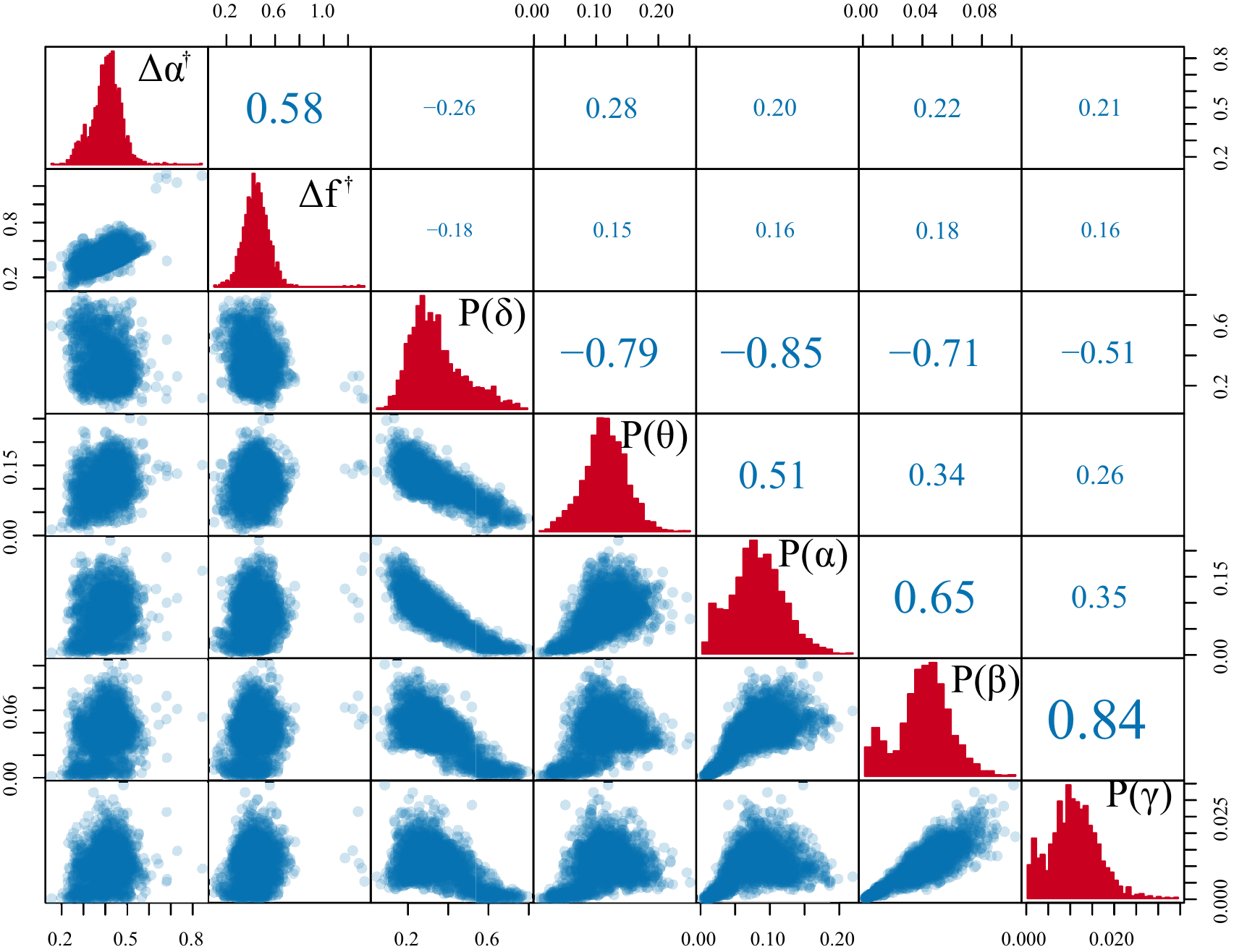}
\caption{{\bf Comparison of multifractal measures with classical spectral band power.}  Scatter plot matrix comparing both standardised multifractal spectrum width and height ($\Delta\alpha^{\dagger}$ and $\Delta f^{\dagger}$) with the $\delta$, $\theta$, $\alpha$, $\beta$, and $\gamma$ average band power in each epoch. Each scatter point is derived from a single epoch of the time series. The diagonal of the matrix features the histograms for each measure. The lower triangle contains the scatter plots for each pair of measures. The upper triangle shows the Pearson correlation for each pair of measure, where the size of the font additionally corresponds to the correlation coefficient to provide an additional visual cue.}
\label{fig:freqBand}
\end{figure}

\subsection*{Experiment 4: Impact of sampling frequency and epoch length on multifractal estimation of human EEG}

The variation of the multifractal spectrum width $\Delta\alpha^{\dagger}$ for different combinations of epoch sizes and sampling frequencies is shown in Fig 7 (A). On visual inspection, it is clear that there are some combinations of epoch size and sampling frequency that show a clear increase of $\Delta\alpha^{\dagger}$ during the ictal period (marked by the red lines). To quantify this effect, Fig 7 (B) shows the Cohen's effect size $D$ of the ictal vs. interictal $\Delta\alpha^{\dagger}$ distributions plotted against epoch duration (in seconds). In this plot, we included 15 different sampling frequencies, and also data from three different EEG channels (all in the seizure onset zone). A peak in $D$ can be seen at about 1 second (across all sampling frequencies), indicating that the change in $\Delta\alpha^{\dagger}$ during a seizure can be best captured when using one second epochs (regardless of sampling frequency). This effect was not found for the sampling frequency or epoch length separately. Similar results for additional patients are shown on Supplementary materials (S2 Appendix).

\begin{figure}[H]
\includegraphics[width=0.9\textwidth]{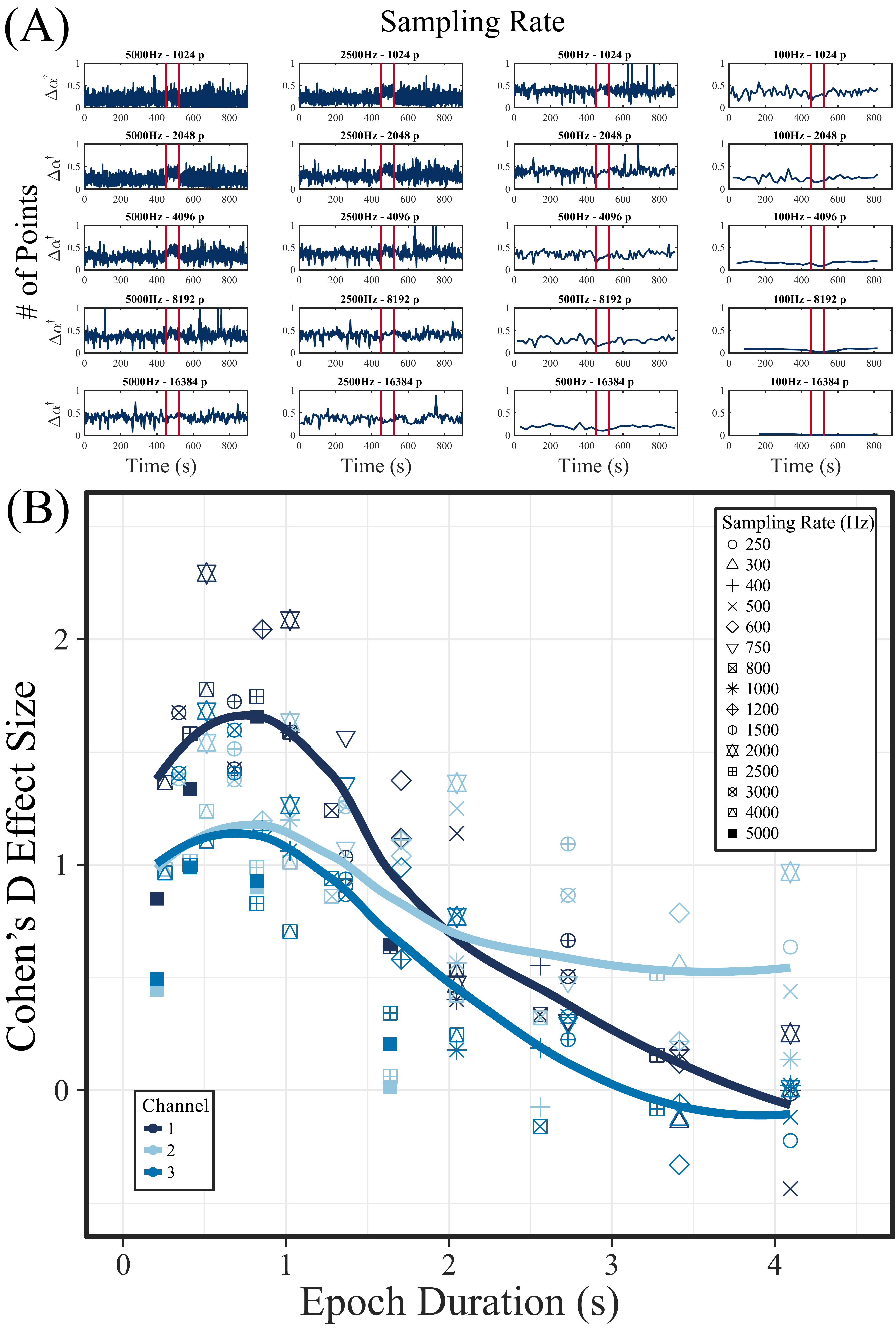}
\caption{{\bf Influence of EEG sampling frequency and epoch length on multifractal spectrum width around and during an epileptic seizure.} (A) Multifractal spectrum width ($\Delta\alpha^{\dagger}$) in a 15-minutes intracranial EEG segment containing one seizure (onset and offset marked by the red lines).  The signal was initially sampled at 5000 Hz. Each column shows $\Delta\alpha^{\dagger}$ for 5000 Hz, 2500 Hz, 500 Hz and 100 Hz sampling rates. Different epoch sizes were used ranging from 1024 to 16384 samples (in each row). (B) Relationship of effect size $D$ (between the interictal and ictal distribution of $\Delta\alpha^{\dagger}$) and epoch duration in seconds (obtained by dividing the number of sampling points by the sampling rate of the signal). Channel 1 is the data shown in (A). The solid line represents a LOESS curve fitting of the data points, with formula ‘$y \sim x$’. The data used for this figure is obtained from for subject 'I001\textunderscore P005\textunderscore D01' around seizure 1. Channel 1: ADMacro\textunderscore 01. Channel 2: ADMacro\textunderscore 02. Channel 3: ADMacro\textunderscore 03.}
\label{fig:winFreq}
\end{figure}

\section*{Discussion}

In this study, we have explored the monofractal and multifractal properties of human EEG recordings and used simulated data to test the performance of fractal property estimation methods. Although mono- and multi-fractal  approaches have been widely employed in the study of physiological signals in humans \cite{Stanley1999,Costa2017,Franca2017,Hu2004,Hu2009,
Ivanov1999}, we have demonstrated that the monofractal dimension may be capturing similar information to the signal variance. When using standardisation to remove the effect of signal variance, we demonstrated that multifractal measures (estimated by the Chhabra-Jensen method) capture information not contained in widely used conventional signal measures.  Finally, using epileptic seizure as an example, we showed that the multifractal estimation epoch length can significantly impact the detection of time-varying effects in multifractal properties, suggesting the need for data- and application-specific optimisation. \\

\subsection*{Methodological considerations}

One of our key observations is that monofractal estimators are still driven by signal variance -- even following epoch-wise standardisation, whereas multifractal properties following epoch-wise standardisation are no longer impacted by signal variance. This is a curious and non-intuitive observation that, to our knowledge, has not been reported before. To explain this observation, it may be worth noting the relationship between monofractal and multifractal analysis. Essentially, in multifractal analysis, at the point for which $q=2$, the $f(\alpha)$ is the so-called correlation dimension, which is an alternative way of estimating the monofractal dimension \cite{Murcio2015}. The multifractal spectral width, however, is the range of values of $\alpha$ that exist for a given signal, and independent of a particular $q$, or its $f(\alpha)$. Hence, conceptually the mono- and multifractal properties we are investigating here are describing different properties of the signal and cannot be directly compared. Based on our results, we hypothesise that the monofractal dimension is not affected by our standardisation method, but that the range of values of $\alpha$, i.e. $\Delta\alpha$ is. Future work has to show (on theoretical grounds) the exact reason for this observation.

We observed that the Chhabra-Jensen method is the most reliable out of the three methods. As was pointed out in the original publication \cite{Chhabra1989}, this is most likely due to the fact that the Chhabra-Jensen method avoids a Legendre transform that the other methods require. The Legendre transformation requires smoothing of the $D_{q}$ curve and can lead to errors. For further advantages of the Chhabra-Jensen method, the reader is referred to the original publication \cite{Chhabra1989}.

%Once the effect of the variance of the signal is removed, it possible to explore and evaluate other properties of the signal. We have shown that $\Delta\alpha^{\dagger}$ is capable of bypassing variance induced changes and reflects physiologic properties in both intracranial and scalp EGG. The two classes of signal were recorded in very different set-ups and with different purposes, with the former an epilepsy surgery planning procedure and the later a polysomnography study (Suppl. S3 Appendix). Such results suggest that the proposed approach (standardisation and sigmoid transform followed by the Chhabra-Jensen method) is generalisable to different classes of signal. Nonetheless, future work should focus on how general the presented properties are, to understand exactly what type of information $\Delta\alpha^{\dagger}$ contains and how it can be exploited for neuroscience research. \\

Finally, our analysis highlighted the importance of choosing an adequate epoch size given a sampling frequency, in order to study events such as epileptic seizures. However, our study was based on the analysis of ictal vs. interictal epochs, i.e. a hard separation that may not represent continuous phenomena accurately. Future work should take into account that multifractal properties may be continuously changing over time (a striking example is shown in Fig. S7), and an explicitly time based approach may be needed. Along similar lines, our finding of a optimal time scale may be due to the non-stationary nature of the multifractal properties. Further theoretical work may have to develop a temporally resolved multifractal estimator, in order to fully understand this aspect.\\

\subsection*{Implications for the understanding of brain activity and brain generators}

Previous studies reported that the brain is characterised by critical dynamics \cite{Eguiluz2005,Chialvo2010,Chialvo2012}. This characteristic, found from microscopic spatial scales (such as neuronal networks) \cite{Beggs2003,Beggs2004} to the global brain structure \cite{Eguiluz2005}, is thought to facilitate the storage and processing of information. It has been further suggested that more than one scaling exponent would be necessary to properly characterise the brain's critical dynamics \cite{Fraiman2012,Papo2014,Papo2017,Suckling2008,
Ciuciu2012,Ihlen2010,Zhang2015,Zorick2013}, as departures from the power-law pattern have been frequently observed in brain signals. Hence, it has been proposed that using additional, higher-order statistical moments can better characterise such data \cite{Fraiman2012}. In this work, we contribute a complementary observation: while monofractal characterisation of EEG appears to essentially follow the slow changes of signal variance, multifractal characterisation is capable of revealing new information. \\

In terms of generative processes that can produce monofractal properties, it has been suggested that a property called Self-Organised Criticality (SOC) \cite{Bak1987} may play an essential role. SOC describes the capacity of a system to evolve naturally into a critical state (a state in which a minimum perturbation could lead to events of all sizes). Such phenomena display power-law distributions and fractal properties as signatures \cite{Bak1995}. An example process that displays SOC is the so-called single avalanche or Bak–Tang–Wiesenfeld model (also known as Abelian sandpile model) \cite{Bak1987}. SOC behaviour has been linked to physiological control mechanisms, such as in human heart rate variability \cite{Goldberger2002}. Similar to SOC, a related regime -- termed non-classical SOC -- is thought to give rise to multifractal properties \cite{Lovejoy2007}. The analysis and understanding of the non-classical SOC is, however, still under development. \\

In this context, our multifractal spectral analyses of human EEG data suggest that cerebral phenomena should not be modelled by a single avalanche model (classical SOC), in agreement with findings in a previous study \cite{Fraiman2012}. Moreover, it is hypothesised that brain dynamics are non-ergodic \cite{Bianco2007}, i.e., display preferential states and depends on previous states \cite{Papo2014}, which are all properties of multifractal processes \cite{Lovejoy2007}. Thus, multifractal analyses could provide a new paradigm for studying brain function and structure, as previously suggested in other studies of normal \cite{Suckling2008,Papo2014,Ciuciu2012,Ihlen2010,
Zorick2013,Papo2017} and pathological brain activity \cite{Zhang2015}.  Furthermore, generative processes displaying multifractal properties could help understanding the observed multifractal changes on a mechanistic level.

\subsection*{On the detection of brain state transitions in health and disease}

A fundamental observation in our work is that an optimal time scales may exist for specific physiological processes (such as epileptic seizures) in terms of their multifractal dynamics (Fig. 7 and supplementary material S2 App.). This result suggests that, at least in an epoch-based study, for any given epileptic seizure in a given patient, the variety of scaling exponents ($\Delta\alpha$) will depend on the length of the epoch analysed. The implications of this observation are that certain scaling exponents will only exist in specific time scales and the diversity of scaling exponents will depend on the duration of the epoch. These results suggest the potential need for ‘tuning’, i.e., potentially having to find the characteristic time for every studied phenomenon. If this is indeed the case, a temporally resolved (not epoch-based) multifractal method should be developed in future to adequately characterise brain dynamics. 

Furthermore, the slow temporal changes in multifractal dynamics need to be characterised in a systematic way. Using epileptic seizures as an example, S7 Fig shows that dramatic changes in multifractal properties can sometimes be seen before an epileptic seizure. This observation requires further investigation to address questions such as: are all epileptic seizures characterised by pre-ictal changes in multifractal properties? Do other physiological processes, such as sleep, influence this finding? \\

\subsection*{Outlook}
Our work has highlighted several challenges that need to be considered when analysing multifractal properties of EEG signals; namely choice of the appropriate estimation method, estimation parameters, and the influence of the signal variance. We have suggested some solutions to these problems, such as the used of the Chhabra-Jensen approach combined with an epoch-wise standardisation approach. We have also highlighted possible process-specific challenges. In terms of epileptic seizures, future work is required to analyse a larger number of patients in order to draw firmer conclusions on the potential clinical relevance of multifractal analyses. Furthermore, the study of mechanistic  generative models of EEG may shed light on why those multifractal changes occur. For example, a generative process of potential interest could feature a modified version of Bak–Tang–Wiesenfeld model \cite{Bak1987}. \\

\subsection*{Summary}
In this paper, we have analysed the monofractal and multifractal properties of human EEG recordings. We have shown that monofractal estimates are influenced by the standard variation of the time series, thus not capturing information beyond signal variance. For multifractal estimation, we have shown that the Chhabra-Jensen approach is the most stable, and we have developed a method of signal pre-processing to remove the influence caused by the variance of the signal. Using the suggested approach, the multifractal estimates do not correlate with traditional EEG measures, thus yielding additional information about the signal. Finally, our results also indicate a preferential time scale to identify differences in multifractal properties between ictal and interictal state recordings in patients with epilepsy.

\section*{Supporting information}

% Include only the SI item label in the paragraph heading. Use the \nameref{label} command to cite SI items in the text.
%\begin{figure}[ht]
%\includegraphics[width=\textwidth]{figs/S1_fig.pdf}
%\end{figure}
\paragraph*{S1 Fig.}
\label{S1_Fig}
{\bf Flow-chart of the Chhabra-Jensen algorithm.} 

\paragraph*{S2 Fig.}
\label{S2_Fig}
{\bf Impact of sigmoid transformation on the signal.} This function maps the original (raw) time series into a sequence of elements ranging in between 0 and 1, allowing the use of the Chhabra-Jensen method. (A) Sigmoid curves with different values of $v$. This parameter defines how shallow/steep the sigmoid curve will be. (B) Effect of the sigmoid transform on an epoch extracted from an intracranial EEG recording. Smaller values of the parameter $v$ tend to flatten the curve. 

\paragraph*{S3 Fig.}
\label{S3_Fig}
{\bf Assessment of parameters for sigmoid mapping function.} (A) Variability of $\Delta\alpha^{\dagger}$ (0.1, 0.5 and 0.9 quantiles) as a function of the parameter $v$. (B) Pearson correlation of the original series and the mapped one for the three types of signal (intracranial EEG, its surrogate and a generated random time series). 

\paragraph*{S4 Fig.}
\label{S4_Fig}
{\bf Monofractal Analysis of seizure segment.} Both Higuchi and DFA approaches exhibits similar patterns to the standard deviation of the signal, in accordance with the simulated data presented on the main text of this article.

\paragraph*{S5 Fig.}
\label{S5_Fig}
{\bf Multifractal spectrum height and other signal property changes over time.} (A) A single channel intracranial EEG time series segment containing one seizure in the patient NHNN1 (channel 1). The seizure onset and offset are marked by red lines. (B) Variation of multifractal spectrum height ($\Delta\alpha$) estimated on epochs of the EEG segment in (A).  (C) Epoch-wise normalised multifractal spectrum height ($\Delta\alpha^{\dagger}$). (D) Standard deviation of the time series for each epoch. (E) Line length of the signal for each epoch.

\paragraph*{S6 Fig.}
\label{S6_Fig}
{\bf Mutual Information of different measures.}  The MI measure shows higher values for the pair $\Delta\alpha$/St. Dev. than to $\Delta\alpha^{\dagger}$/St. Dev., similar to the result in Fig. 5.

\paragraph*{S7 Fig.}
\label{S7_Fig}
{\bf Multifractal spectrum width ($\Delta\alpha^{\dagger}$) for a single icEEG channel in the patient NHNN1 (channel 1) around a seizure.} Black line: moving average.

%\paragraph*{S8 Fig.}
%\label{S8_Fig}
%{\bf Density scatter plots.} (A) $\Delta\alpha$ vs. St. Dev. (B) $\Delta\alpha^{\dagger}$ vs. St. Dev. 

\paragraph*{S1 Appendix.}
\label{S1_Appendix}
{\bf Fractional Brownian motion.}

\paragraph*{S2 Appendix.}
\label{S2_Appendix}
{\bf Characteristic time scale for other iEEG subjects.}

\paragraph*{S3 Appendix.}
\label{S3_Appendix}
{\bf Multifractal properties of sleep EEG.}

\paragraph*{S1 Table.}
{\bf Patient information table.}

%\paragraph*{S2 Fig.}
%\label{S2_Fig}
%{\bf Lorem ipsum.} Maecenas convallis mauris sit amet sem ultrices gravida. Etiam eget sapien nibh. Sed ac ipsum eget enim egestas ullamcorper nec euismod ligula. Curabitur fringilla pulvinar lectus consectetur pellentesque.
%
%\paragraph*{S1 File.}
%\label{S1_File}
%{\bf Lorem ipsum.}  Maecenas convallis mauris sit amet sem ultrices gravida. Etiam eget sapien nibh. Sed ac ipsum eget enim egestas ullamcorper nec euismod ligula. Curabitur fringilla pulvinar lectus consectetur pellentesque.
%
%\paragraph*{S1 Video.}
%\label{S1_Video}
%{\bf Lorem ipsum.}  Maecenas convallis mauris sit amet sem ultrices gravida. Etiam eget sapien nibh. Sed ac ipsum eget enim egestas ullamcorper nec euismod ligula. Curabitur fringilla pulvinar lectus consectetur pellentesque.
%
%\paragraph*{S1 Table.}
%{\bf P-values Kruskal Wallis test for multifractal spectra width and height, and $\delta$-band power in different sleep stages.}

\section*{Acknowledgments}
LGSF was supported by a grant from Brazilian Brazilian National Council for Scientific and Technological Development (CNPq) (206907/2014-1). The authors acknowledge the use of the UCL Legion High Performance Computing Facility (Legion@UCL), and associated support services, in the completion of this work. \\

The authors would like to thank Benjamin H. Brinkmann, Joost Wagenaar, and Hoameng Ung from iEEG.org.

\section*{Author contributions}

\textbf{Conceptualisation:} LGSF, LL, MCW, YW\\
\textbf{Data Curation:} LGSF\\
\textbf{Formal analysis:} LGSF, YW\\
\textbf{Funding acquisition:} LGSF\\
\textbf{Investigation:} LGSF\\
\textbf{Methodology:} JGVM, LGSF, LL, MCW, YW\\
\textbf{Project Admnistration:} LGSF, LL, MCW, YW\\
\textbf{Resources:} LGSF, LL, MCW, YW\\
\textbf{Software} JGVM, LGSF, YW\\
\textbf{Supervision:} LL, MCW, YW\\
\textbf{Validation:} LGSF, YW\\
\textbf{Visualisation:} LGSF\\
\textbf{Writing – original draft:} LGSF, LL, MCW, YW\\
\textbf{Writing – review \& editing:} JGVM, LGSF, LL, MCW, ML,  NKS, YW

% Either type in your references using
% \begin{thebibliography}{}
% Text
% \end{thebibliography}
%
% or
%
% Compile your BiBTeX database using our plos2015.bst
% style file and paste the contents of your .bbl file
% here. See http://journals.plos.org/plosone/s/latex for 
%\begin{thebibliography}{10}
%
%\bibitem{bib1}
%Conant GC, Wolfe KH.
%\newblock {{T}urning a hobby into a job: how duplicated genes find new
%  functions}.
%\newblock Nat Rev Genet. 2008 Dec;9(12):938--950.
%
%\bibitem{bib2}
%Ohno S.
%\newblock Evolution by gene duplication.
%\newblock London: George Alien \& Unwin Ltd. Berlin, Heidelberg and New York:
%  Springer-Verlag.; 1970.
%
%\bibitem{bib3}
%Magwire MM, Bayer F, Webster CL, Cao C, Jiggins FM.
%\newblock {{S}uccessive increases in the resistance of {D}rosophila to viral
%  infection through a transposon insertion followed by a {D}uplication}.
%\newblock PLoS Genet. 2011 Oct;7(10):e1002337.

%\end{thebibliography}
%\documentclass[10pt,a4paper]{article}
%\usepackage[utf8]{inputenc}
%\usepackage{amsmath}
%\usepackage{amsfonts}
%\usepackage{amssymb}
%\usepackage{url}
%\usepackage[left=2cm,right=2cm,top=2cm,bottom=2cm]{geometry}
%
%\bibliographystyle{plos2015_updated}
%
%\begin{document}

\section*{S1 Appendix: Fractional Brownian Motion}

A fractional Brownian motion is a continuous zero-mean Gaussian process that can be described by the function in equation \ref{eq:fBm} \cite{Kroese2015}. In order to simulate such properties, it is sufficient to simulate the increments on the profile, as described in equation \ref{eq:increments}.

\begin{equation}
\xi_{1} = B_{1}^H, \xi_{2} = B_{2}^H - B_{1}^H, \xi_{N} = B_{N}^H - B_{N-1}^H
\label{eq:increments}
\end{equation}

These values form a stationary sequence of standard Gaussian variables with a covariance described in equation \ref{eq:fBm}. The vector $\xi = (\xi_{1},...,\xi_{N})^{T}$, composed by the simulated increments, is called fractional Gaussian noise and is a centred Gaussian array with covariance matrix $\Omega$ \cite{Shevchenko2014}.

\begin{equation}
\rho_{H}(n) = E[\xi_{1}\xi_{n+1}] = \dfrac{1}{2}\bigg((n+1)^{2H} + (n-1)^{2H} - 2n^{2H}\bigg), \quad n \geq 1
\label{eq:fBm}
\end{equation}

\[
\Omega = \text{Cov}(\xi) =
\begin{pmatrix}
  1 & \rho_{H}(1) & \rho_{H}(2) & \cdots & \rho_{H}(N-2) & \rho_{H}(N-1) \\
  \rho_{H}(1) & 1 & \rho_{H}(1) & \cdots & \rho_{H}(N-3) & \rho_{H}(N-2) \\
  \rho_{H}(2) & \rho_{H}(1) & 1 & \cdots & \rho_{H}(N-4) & \rho_{H}(N-3) \\
  \vdots & \vdots & \vdots & \ddots & \vdots & \vdots \\
  \rho_{H}(N-2) & \rho_{H}(N-3) & \rho_{H}(N-4) & \cdots & 1 & \rho_{H}(1) \\
  \rho_{H}(N-1) & \rho_{H}(N-2) & \rho_{H}(N-3) & \cdots & \rho_{H}(1) & 1
\end{pmatrix}
\]
\label{eq:covMat}

In order to solve this problem, we can, alternatively, convert the covariance matrix into a circulant matrix. The circular matrix ($\Sigma$) has a known eigenvalues structure, making it easier to solve the problem. Taking the relationship $M = 2(N-1)$, a circular matrix $\Sigma$ is defined based on the relations defined by equation \ref{eq:cRel}.

\begin{equation}
	\begin{aligned}
		c_0 &= 1,\\
		c_{k} &= \begin{cases}
			\rho_{H}(n), \quad n=1,2,...,N-1\\
			\rho_{H}(M-n), \quad n=N,N+1,...,M-1
        \end{cases}
	\end{aligned}
	\label{eq:cRel}
\end{equation}

\[
\Sigma = circ(c_{0},c_{1},...,c_{M-1}) =
\begin{pmatrix}
  c_{0} & c_{1} & c_{2} & \cdots & c_{M-2} & c_{M-1} \\
  c_{M-1} & c_{0} & c_{1} & \cdots & c_{M-3} & c_{M-2} \\
  c_{M-2} & c_{M-1} & c_{0} & \cdots & c_{M-4} & c_{M-3} \\
  \cdots & \cdots & \cdots & \ddots & \cdots & \cdots \\
  c_{2} & c_{3} & c_{4} & \cdots & c_{0} & c_{1} \\
  c_{1} & c_{2} & c_{3} & \cdots & c_{M-1} & c_{0} \\
\end{pmatrix}
\]

We then seek for a factorisation described in the equation \ref{eq:lambda} \cite{Kroese2015}. The increments in the simulated Brownian motion time series will be derived from the eigenvalues $\lambda = (\lambda_{1},...,\lambda_{4N^2})^{T}$ arranged as a $2N \text{x} 2N$ matrix $\Lambda$. The variable $P$ is the Kronecker product of two discreet Fourier transform matrices given by $P = F \bigotimes F$  \cite{Kroese2015}.

\begin{equation}
\Sigma = P^* \Lambda P
\label{eq:lambda}
\end{equation}

\begin{equation}
F_{jk} = \dfrac{1}{\sqrt{2N}}e^{-2\pi ijk/2N}, \quad \text{where } \quad j,k=0,1,...2N-1
\end{equation}

The algorithm used to derive the fBm time series is available at \url{https://github.com/lucasfr/ModfBm} and can be summarised in the following steps: First, we compute the covariance matrix ($\Omega$)(step 1) and build a $2N \text{x } 2N$ circulant matrix ($\Sigma$) (step 2). For this operation, we just need the first row of the matrix $\Sigma$ (step 3). The eigenvalues of matrix $\Lambda$ can then be obtained via Fourier transformation. Both real and complex parts will present a Gaussian profile, however, it is enough to consider the real part of the eigenvalues (step 4).\\
The following step consists of multiplying the output of step 4 by random complex numbers and applying the inverse Fourier transformation (step 5). We modified this step and added an element of a function M that will be multiplied by these complex random values before applying the inverse Fourier transformation. From now on, this modified version will be denoted as Modulated fractional Brownian motion (ModfBm)\\
The last step step consists of accumulating the values generated by the previous operation and multiplying by a constant scaled by the the Hurst exponent $H$. The generated time series should present scale properties according to the chosen $H$ exponent and, for our ModfBm, feature changes on its standard deviation.\\

We generated a time series of 1.843.200 points (1800 windows of 1024 points), corresponding to a recording of an hour of duration with a sampling rate of 512 Hz - similar to the clinical intracranial EEG segment from the patient 'NHNN1'. The time series was simulated with a Hurst exponent $H = 0.7$, the value was chosen due to its persistent features, i.e., it generates a time series with memory. Additionally, a modulating function $M$, for every 2 seconds window $w$, described by \ref{eq:Mfunction} and shown in Fig. 3(C) was used to simulate the Modulated fBm.

\begin{equation}
		M(w) =
		\begin{cases}
			1, \quad w < 450\\
			1 + (w-449)0.01, \quad 450 \leq w \leq 900\\
			1, \quad w > 900
        \end{cases}
        \label{eq:Mfunction}
\end{equation}

%\bibliography{ref}
%
%\end{document}
\clearpage

%\documentclass[10pt,a4paper]{article}
%\usepackage[utf8]{inputenc}
%\usepackage{amsmath}
%\usepackage{amsfonts}
%\usepackage{amssymb}
%\usepackage{url}
%\usepackage[left=2cm,right=2cm,top=2cm,bottom=2cm]{geometry}
%\usepackage{graphicx}
%\usepackage{caption}
%
%\begin{document}

\section*{S2 Appendix: Characteristic time scale for other iEEG subjects}
%how does the original time series look like, which  patient were they, where were the channels taken from, etc.
All the following datasets analysed are taken from the iEEG database (\url{https://www.ieeg.org/}). Study IDs are shown as on the database. The example channels we analysed were labelled seizure onset zone channels. The analysis method is the same as for the main text Fig.~7.

\begin{figure}[H]
\includegraphics[width=\textwidth]{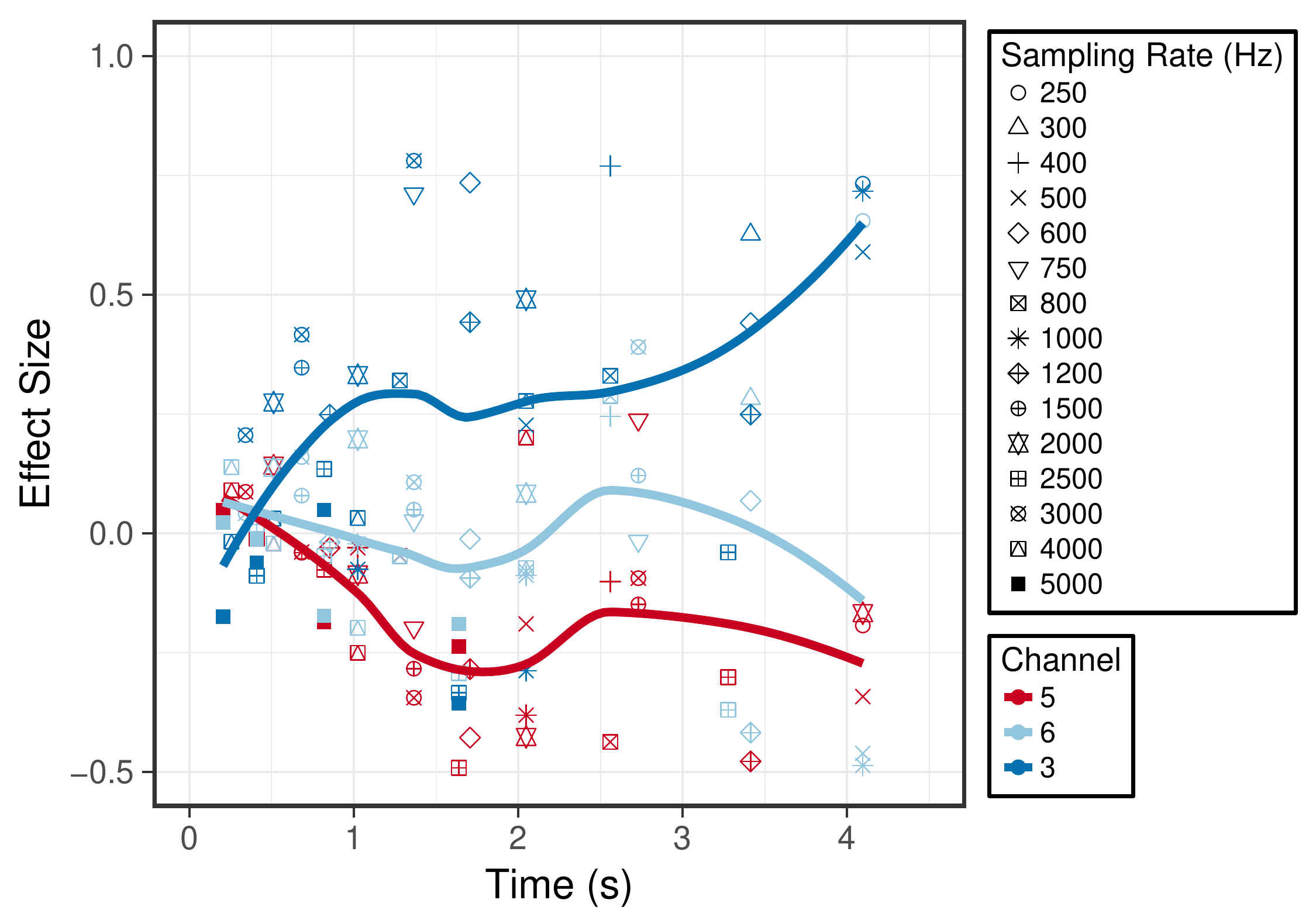}
\caption{{\bf I001\textunderscore P010\textunderscore D01. The three different channels are 'ITS\textunderscore 1','ITS\textunderscore 2', 'LG\textunderscore 08'. } 
}
\label{fig:I001P010D01}
\end{figure}

\begin{figure}[H]
\includegraphics[width=\textwidth]{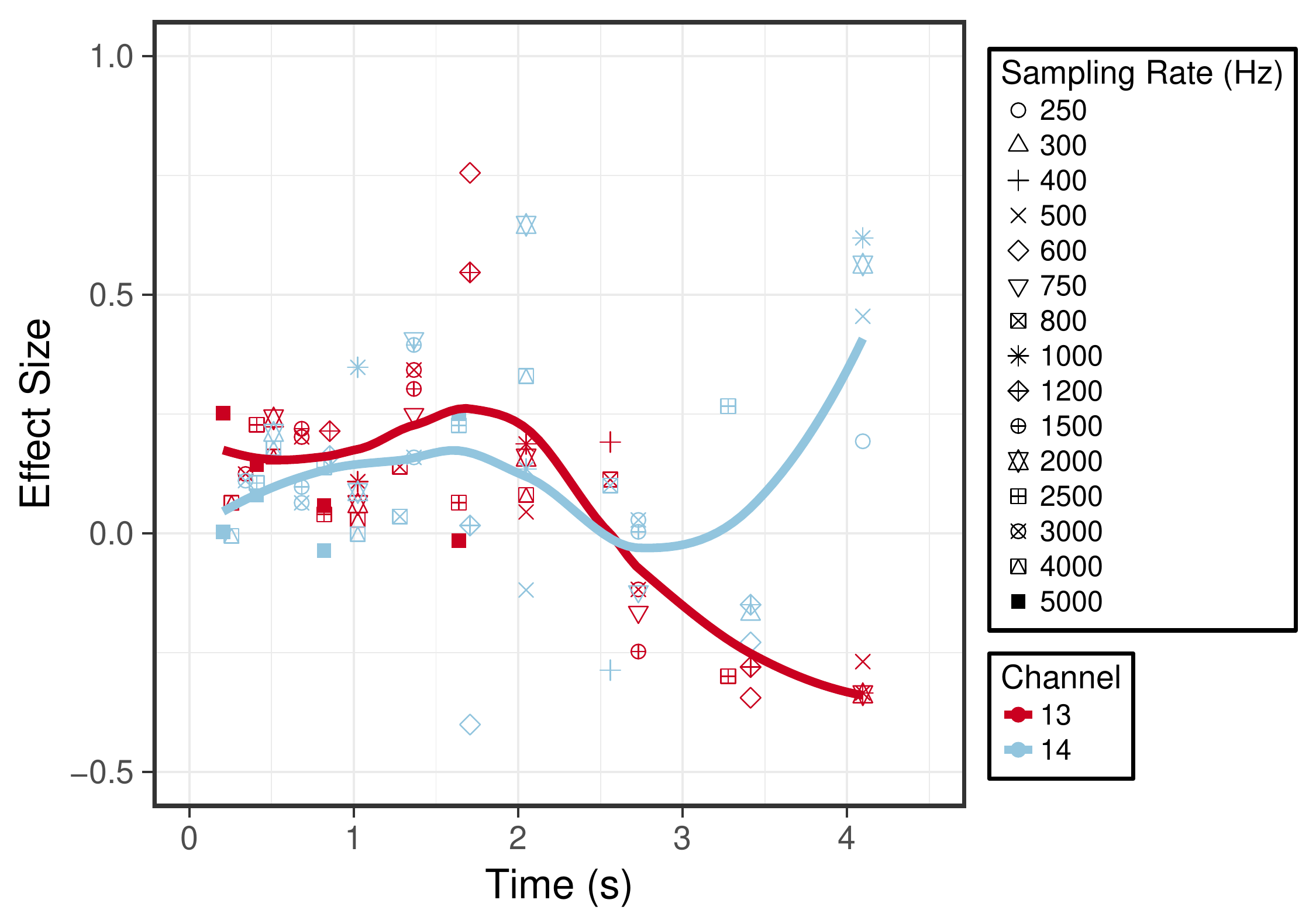}
\caption{{\bf I001\textunderscore P034\textunderscore D01. The two different channels are 'Grid22' and 'Grid23'.} 
}
\label{fig:I001P034D01}
\end{figure}

\begin{figure}[H]
\includegraphics[width=\textwidth]{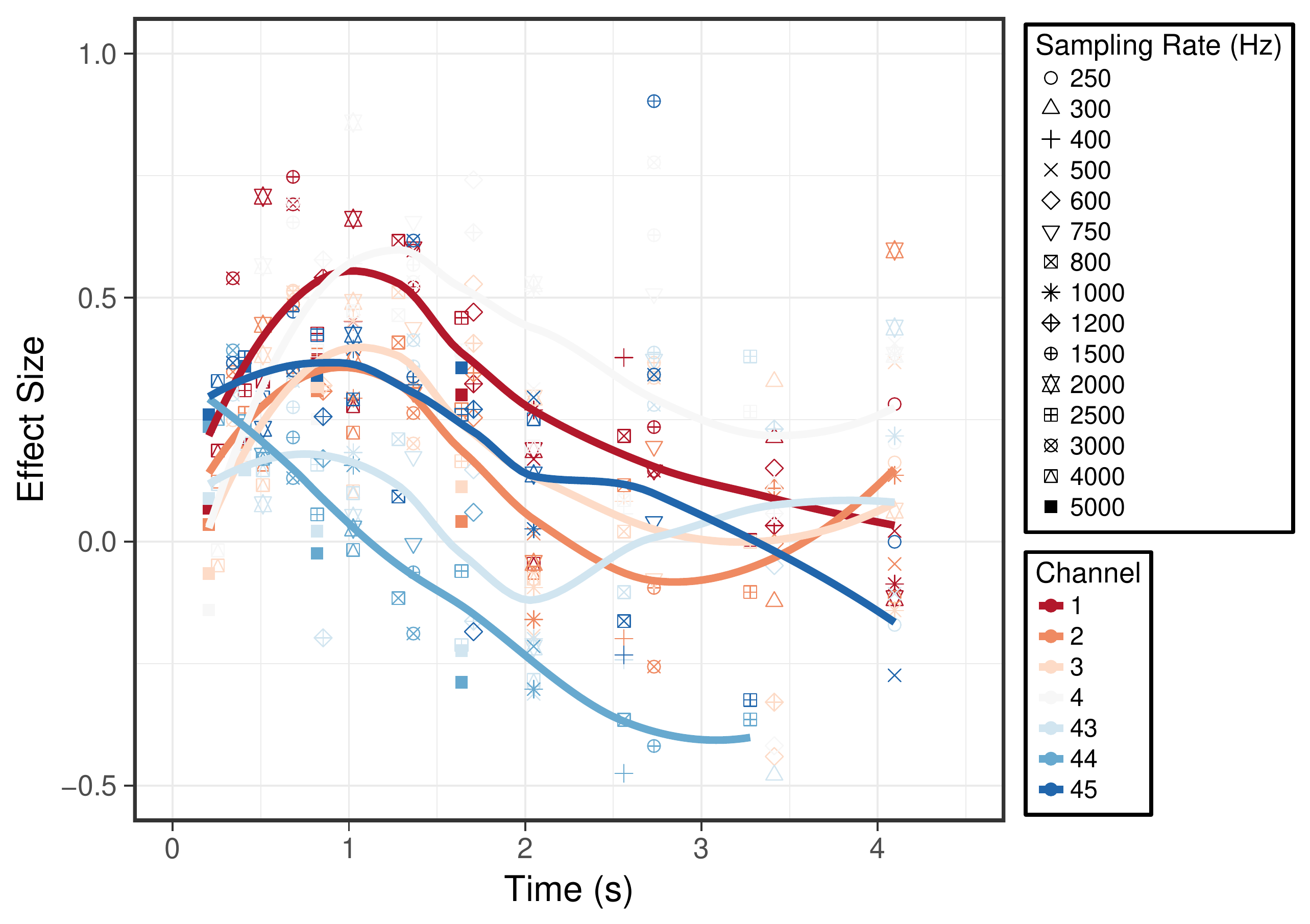}
\caption{{\bf Study 040. The seven different channels are 'LD1', 'LD2', 'LD3', 'LD4', 'LG44', 'LG45', 'LG46'.} 
}
\label{fig:study040}
\end{figure}
%
%\end{document}
\clearpage

%\documentclass[10pt,a4paper]{article}
%\usepackage[utf8]{inputenc}
%\usepackage{amsmath}
%\usepackage{amsfonts}
%\usepackage{amssymb}
%\usepackage{url}
%\usepackage[left=2cm,right=2cm,top=2cm,bottom=2cm]{geometry}
%\usepackage{graphicx}
%\usepackage{caption}
%\bibliographystyle{plos2015_updated}
%
%\begin{document}

\section*{S3 Appendix: Multifractal properties of sleep EEG}

We applied multifractal analysis to scalp EEG obtained from subjects in a polysomnography set-up.  The main goal was to study the impact of the sleep phase on multifractal measures. For this we used the Chhabra-Jensen method on time data that was epoch-wise standardised and sigmoid transformed (i.e. the established pipeline we used in the main manuscript). Each epoch was 10.24 seconds long, and the sampling frequency of the data was 100 Hz giving 1024 points per epoch.

The scalp EEG data studied are from the PhysioNet repository (https://physionet.org/)  [51]\cite{Goldberger2000}  and were described in details by Kemp et al.  [54]\cite{Kemp2000}. This article reports analysis performed on signals of five of these patients: ’ST7011J’, ’ST7022J’, ’ST7041J’, ’ST7052J’, ’ST7061J’. The signals were recorded at 100 Hz in a polysomnography set-up, this study focused on the available EEG recordings only, with two different signals: Fpz-Cz and Pz-Oz.

\begin{figure}[ht!]
\centering
\includegraphics[width=0.9\textwidth]{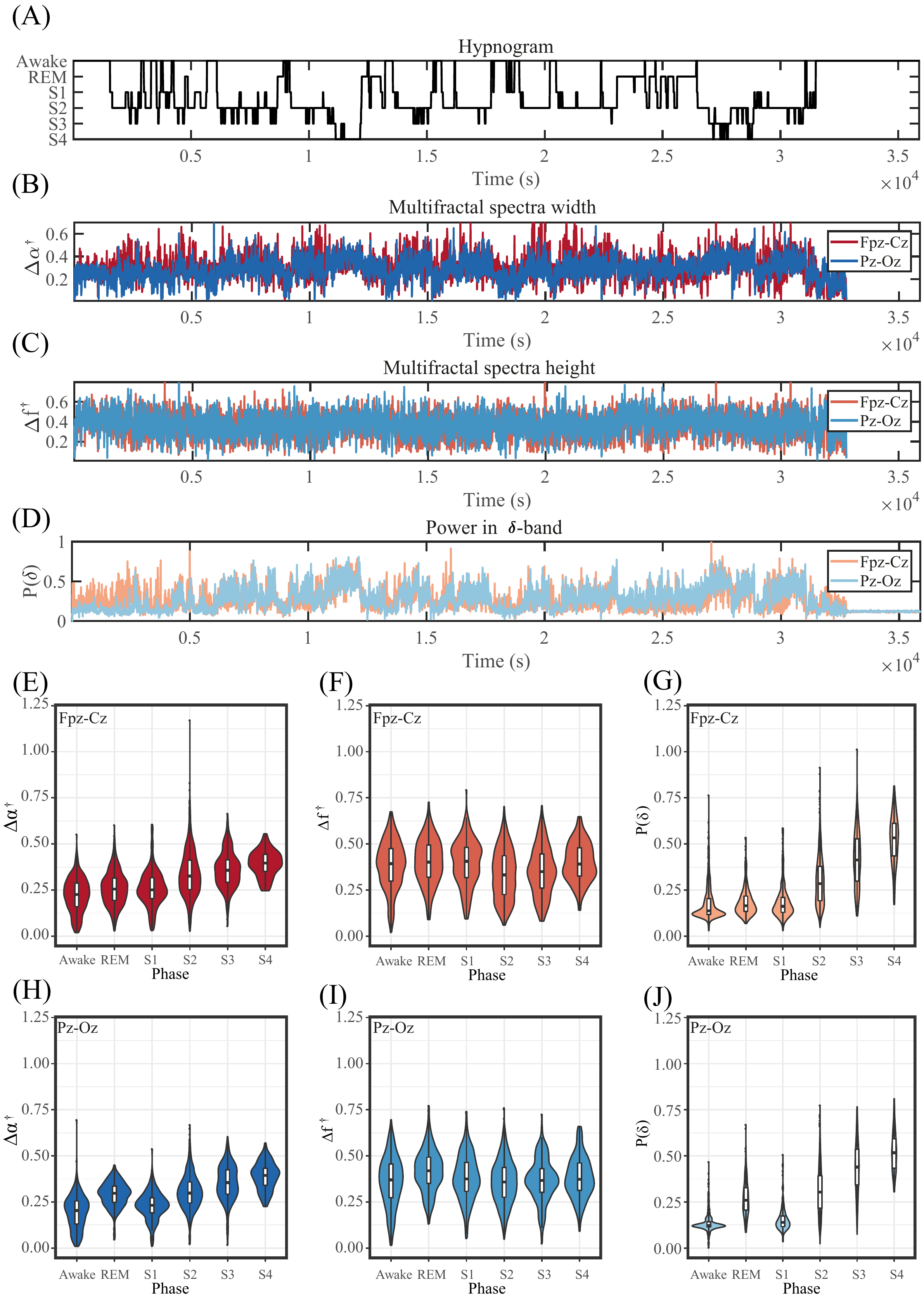}
\caption{{\bf Impact Sleep phases.} The figure features measures performed on a scalp EEG recorded in a polysomnography in two montages Fpz-Cz and Pz-Oz for patient ST7011J. A) Hypnogram with sleep phases marked by a specialist. B) Variation of the Multifractal spectra width in time. C) Variation of the Multifractal spectra height in time. D) Variation of the power of the $\delta$ band in time. E, F, G, H, I, and J) Violin plots of the estimated values of $\Delta\alpha^\dagger$ (E, H), $\Delta f^\dagger$ (F, I), and $P(\delta)$ (G, J) for both Fpz-Cz (E, F, G) and Pz-Oz (H, I, J) montages. At visual inspection, $\Delta\alpha^{\dagger}$ presents a behaviour similar to $P(\delta)$, following the variations of the hypnogram. The violin plots show an increase in the measures towards more advanced sleep phases for both $\Delta\alpha^{\dagger}$ and $P(\delta)$, in both montages.} 
\label{fig:sleep}

\end{figure}

Figure \ref{fig:sleep} shows the variation for $\Delta\alpha^{\dagger}$ (panel B) and $\Delta f^{\dagger}$ (panel C) for two EEG channels: Fpz-Cz and Pz-Oz. %Both channels present similar patterns for the multifractal metrics. The variation of the $\Delta\alpha^{\dagger}$ is also similar to the variation of power in $\delta$ band (slow waves), reaching a correlation of $\rho = 0.60$ in Pz-Oz channel. \\
We also plotted the delta power in Figure~\ref{fig:sleep}(D) as reference, which is often a proxy for slow wave sleep stages. To analyse the relationship between the sleep stages (shown in Figure~\ref{fig:sleep}(A)) and the multifractal measures, we divided the segments according to the sleep stages. A violin plots shows the distribution of  $\Delta\alpha^{\dagger}$, $\Delta f^{\dagger}$, and power in the $\delta$ band (Fig.~\ref{fig:sleep}E-J). A clear drift in the $\Delta\alpha^{\dagger}$ distribution can be observed that slowly changes from awake to REM to S1, S2, S3 and S4 (Fig.~\ref{fig:sleep}E and H). This is in stark contrast to power in the $\delta$ band, where the distributions between awake, REM, and S1 are very similar, and then become highly variable in S2, S3 and S4 (Fig.~\ref{fig:sleep}G and J).

\begin{figure}[ht]
\centering
\includegraphics[width=\textwidth]{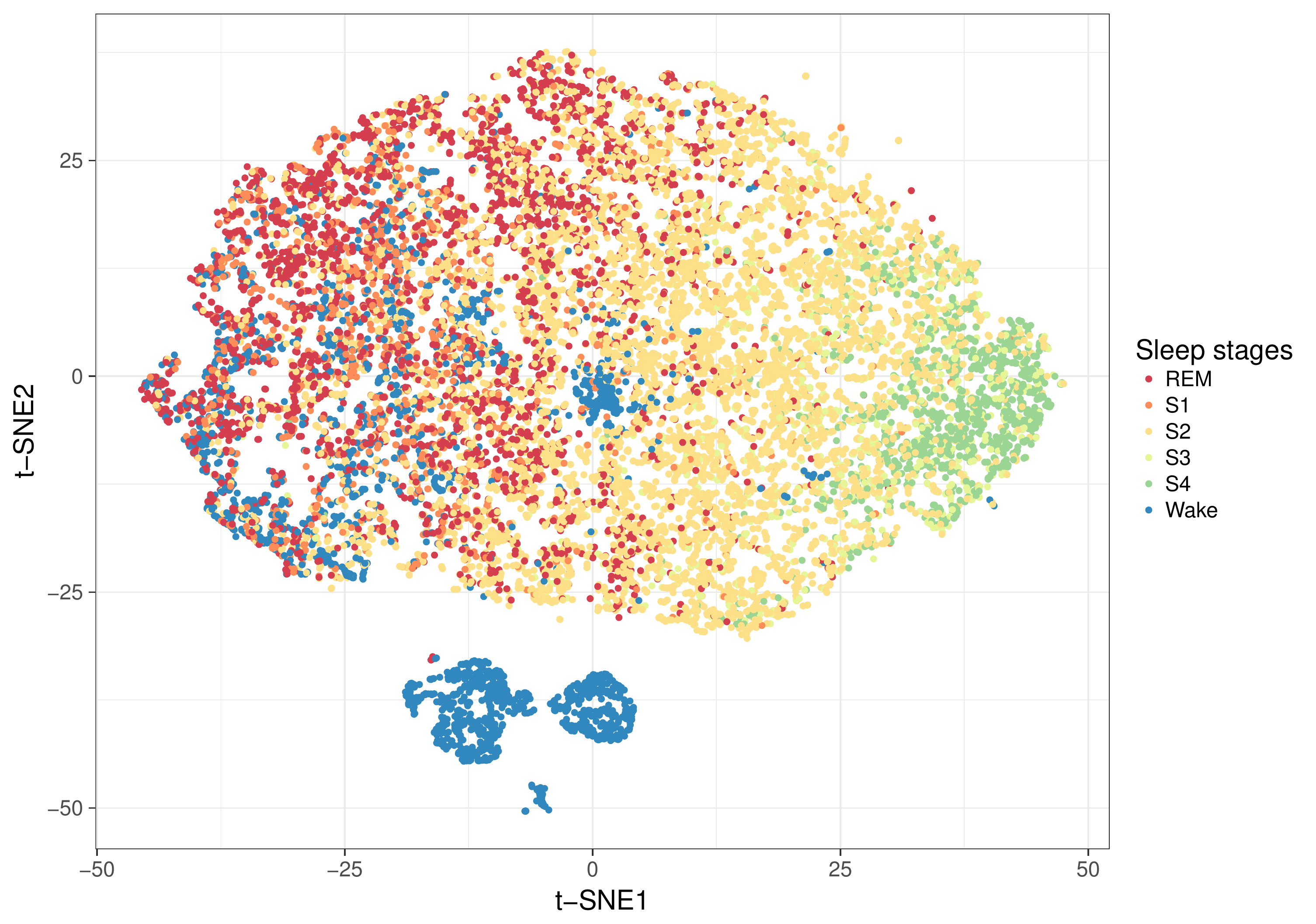}
\caption{{\bf Dimensionality reduction.} t-distributed stochastic neighbor embedding of the $\Delta\alpha^{\dagger}$, $\Delta f^{\dagger}$, and $P(\delta)$ variables for both Fpz-Cz and Pz-Oz channels in a two dimensional space. The figure shows all sleep stages. It is possible to observe that the measures provide information about the the sleep stage that could be used in order to devise a classifier.
}
\label{fig:tSNE}
\end{figure}

A dimensionality reduction technique called t-distributed stochastic neighbour embedding (t-SNE) \cite{VanDerMaaten2008} was applied to the six variable collected from the two channels. Fig \ref{fig:tSNE} shows the projection into a two-dimensional space. Note that the information about the sleep stages were not given to the tSNE algorithm. The projection shows one big cluster and two smaller clusters. When applying the information about the sleep stages, it becomes clear that the two smaller clusters are awake states, whereas the one big cluster shows a progression from REM to S4. Note that there are also datapoints from the awake state in this cluster. Future work will show if those datapoints are simply mislabeled, or in any way closer to sleep states. If so, multifractal measures may be useful in devising a classifier for sleep stages in EEG.

%
%\bibliography{ref}
%
%\end{document}
\clearpage

\section*{S1 Fig}
\begin{figure}[h]
\includegraphics[width=\textwidth]{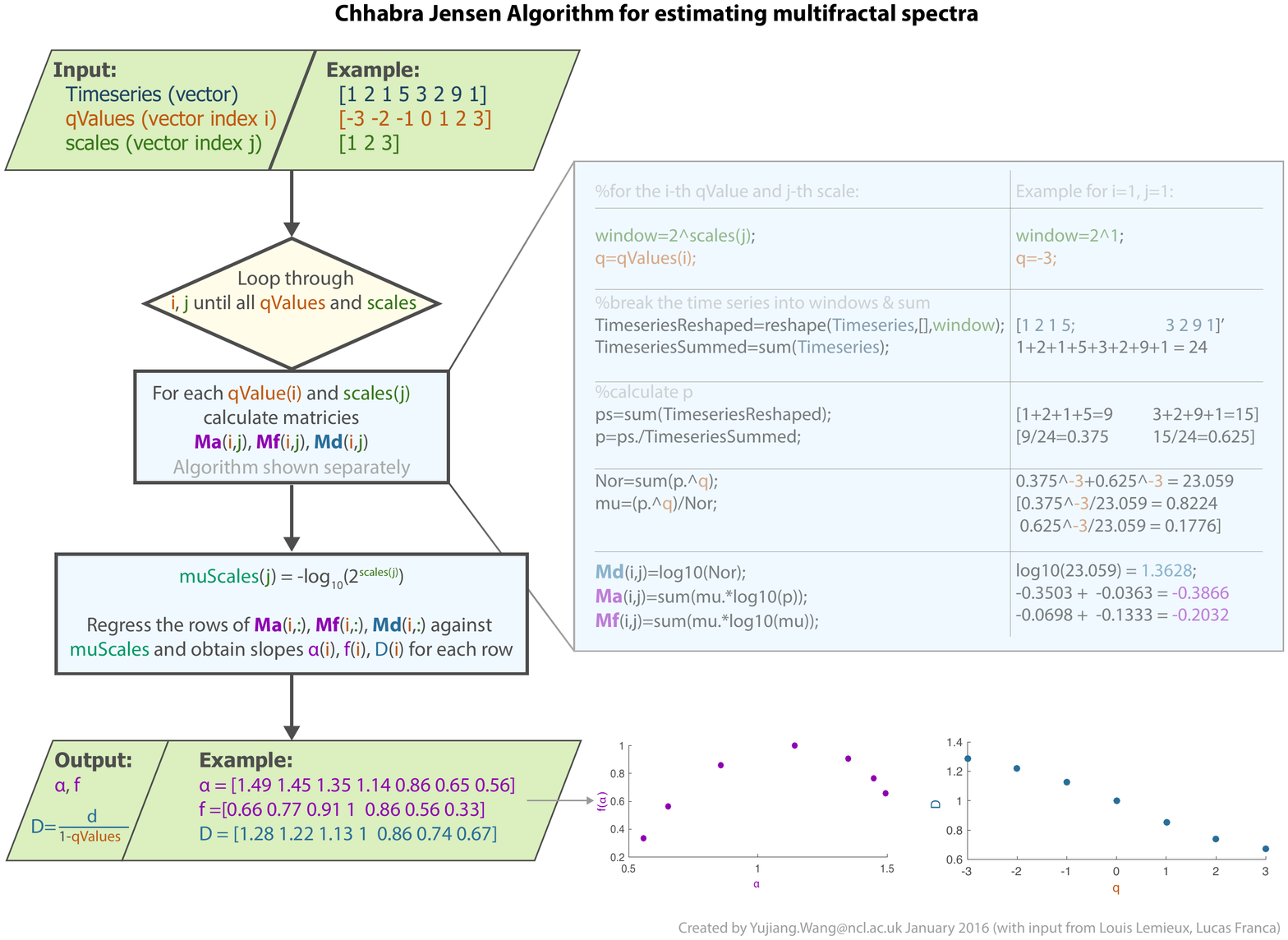}
\caption{{\bf Flow-chart of the Chhabra-Jensen algorithm.} 
}
\end{figure}
\clearpage

\section*{S2 Fig}
\begin{figure}[h]
\includegraphics[width=\textwidth]{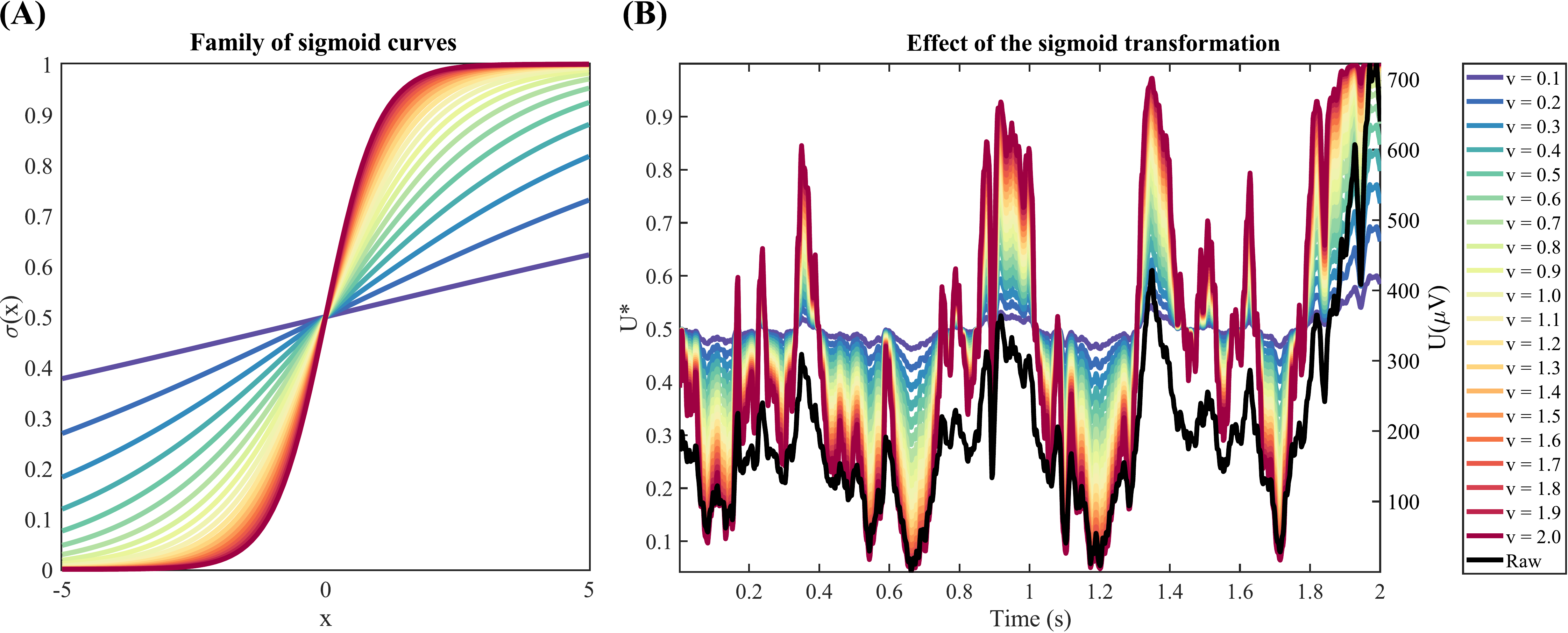}
\caption{{\bf Impact of sigmoid transformation on the signal.} This function maps the original (raw) time series into a sequence of elements ranging in between 0 and 1, allowing the use of the Chhabra-Jensen method in recordings originally containing negative values. (A) Sigmoid curves with different values of $v$. This parameter defines how shallow/steep the sigmoid curve will be. (B) Effect of the sigmoid transform on an epoch extracted from an intracranial EEG recording. Smaller values of the parameter $v$ tend to flatten the curve.  
}
\end{figure}
\clearpage

\section*{S3 Fig}
\begin{figure}[h]
\includegraphics[width=\textwidth]{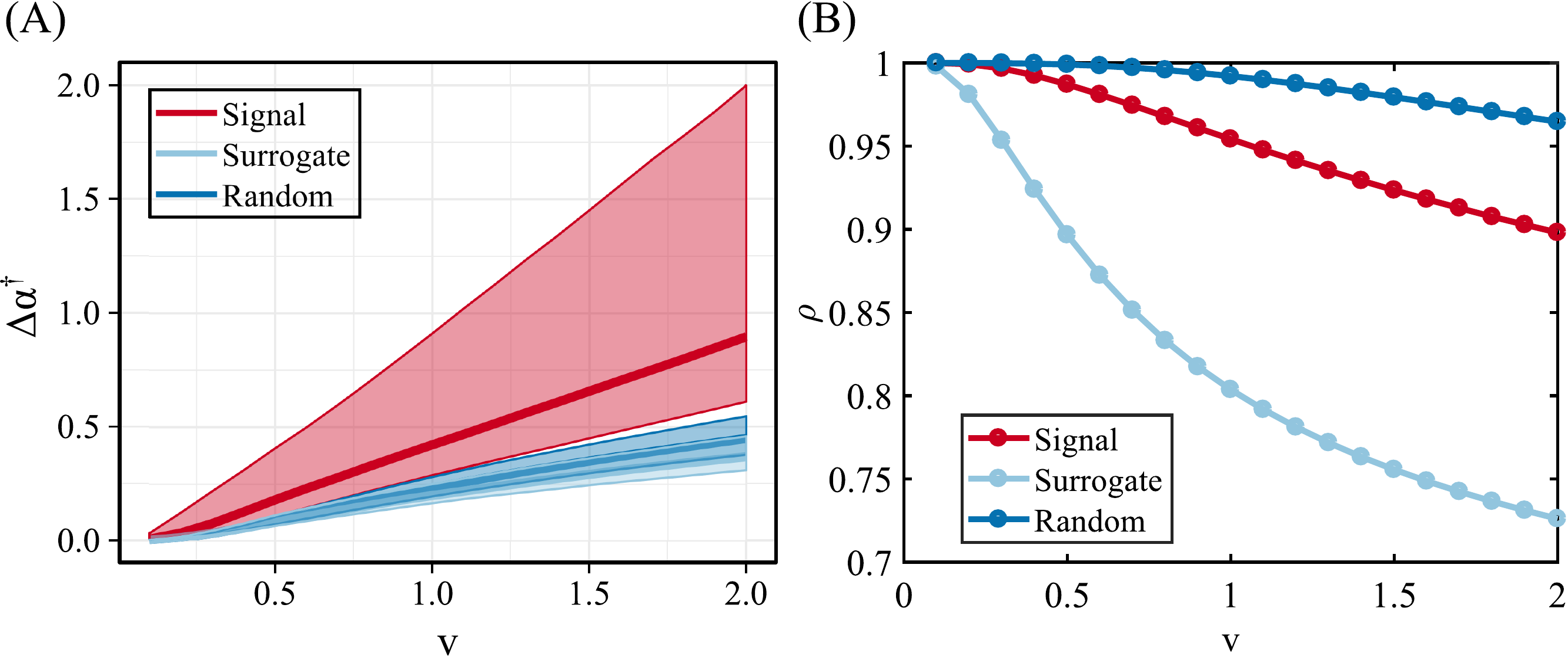}
\caption{{\bf Assessment of parameters for sigmoid mapping function.} (A) Variability of $\Delta\alpha^{\dagger}$ (0.1, 0.5 and 0.9 quantiles) as a function of the parameter $v$. (B) Pearson correlation of the original series and the mapped one for the three types of signal (intracranial EEG, its surrogate and a generated random time series). Based on the optimisation criterion - maximum difference between the different signals and minimum distortion (maximum correlation) – the value chosen was $v=1$. 
}
\end{figure}
\clearpage

\section*{S4 Fig}
\begin{figure}[h]
\centering
\includegraphics[width=0.72\textwidth]{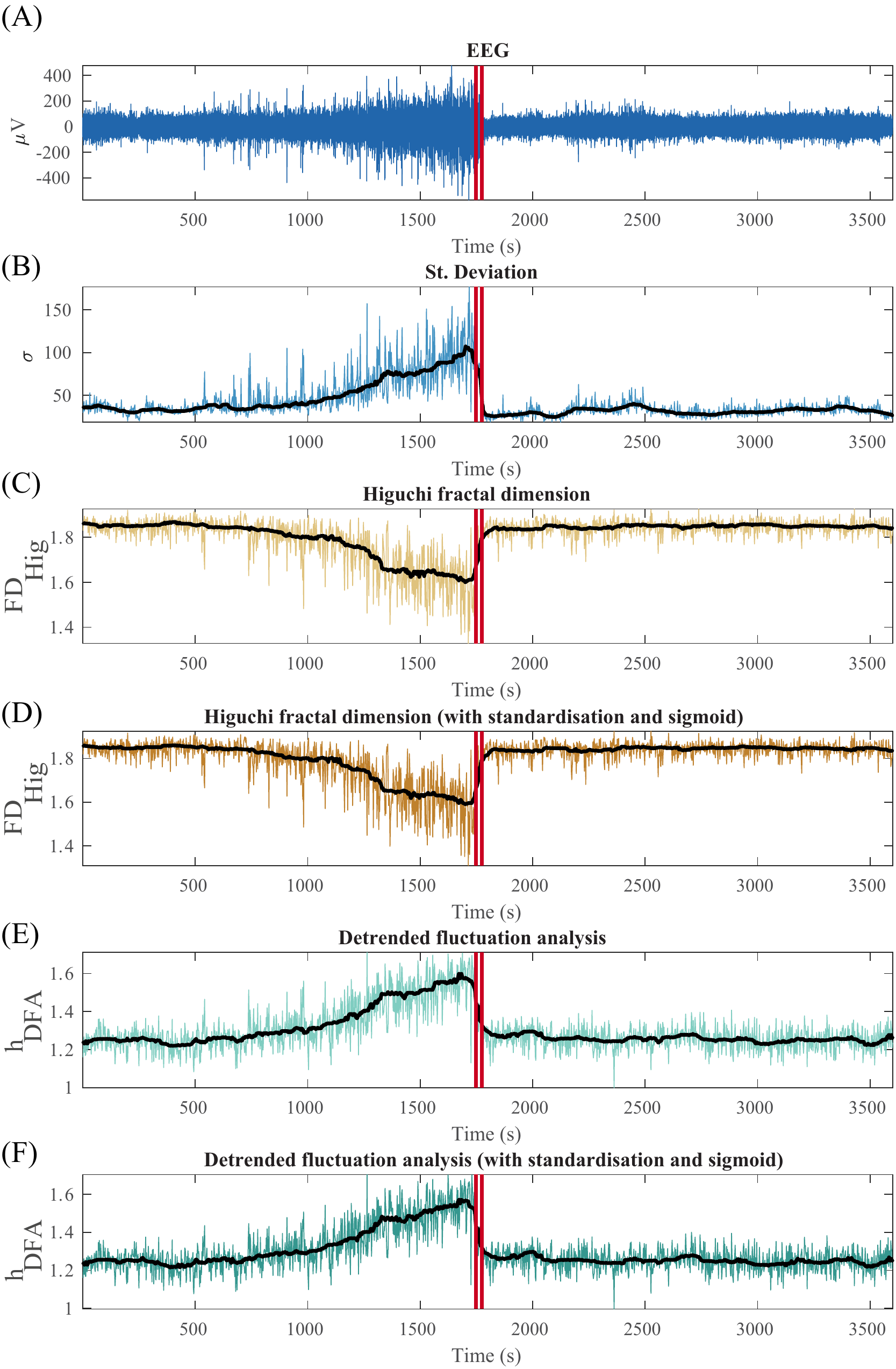}
\caption{{\bf Monofractal Analysis of seizure segment. Both Higuchi and DFA approaches exhibits similar patterns to the standard deviation of the signal, in accordance with the simulated data presented on the main text of this article. 
}}
\end{figure}
\clearpage

\section*{S5 Fig}
\begin{figure}[h]
\includegraphics[width=\textwidth]{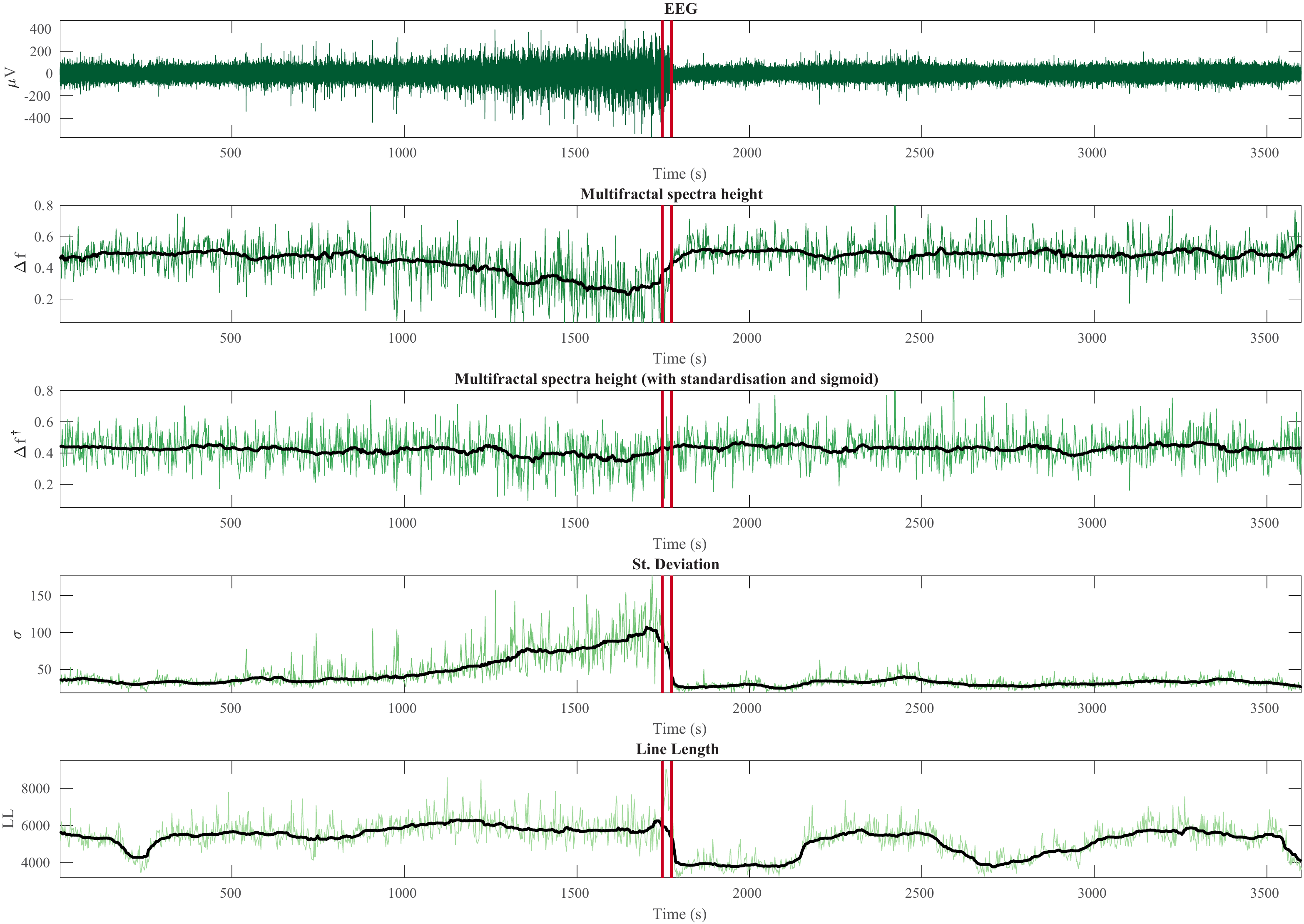}
\caption{{\bf Multifractal spectrum height and other signal property changes over time.} (A) A single channel intracranial EEG time series segment containing one seizure in the patient NHNN1 (channel 1). The seizure onset and offset are marked by red lines. (B) Variation of multifractal spectrum height ($\Delta\alpha$) estimated on epochs of the EEG segment in (A).  (C) Epoch-wise normalised multifractal spectrum height ($\Delta\alpha^{\dagger}$). (D) Standard deviation of the time series for each epoch. (E) Line length of the signal for each epoch.
}
\end{figure}
\clearpage

\section*{S6 Fig}
\begin{figure}[h]
\includegraphics[width=\textwidth]{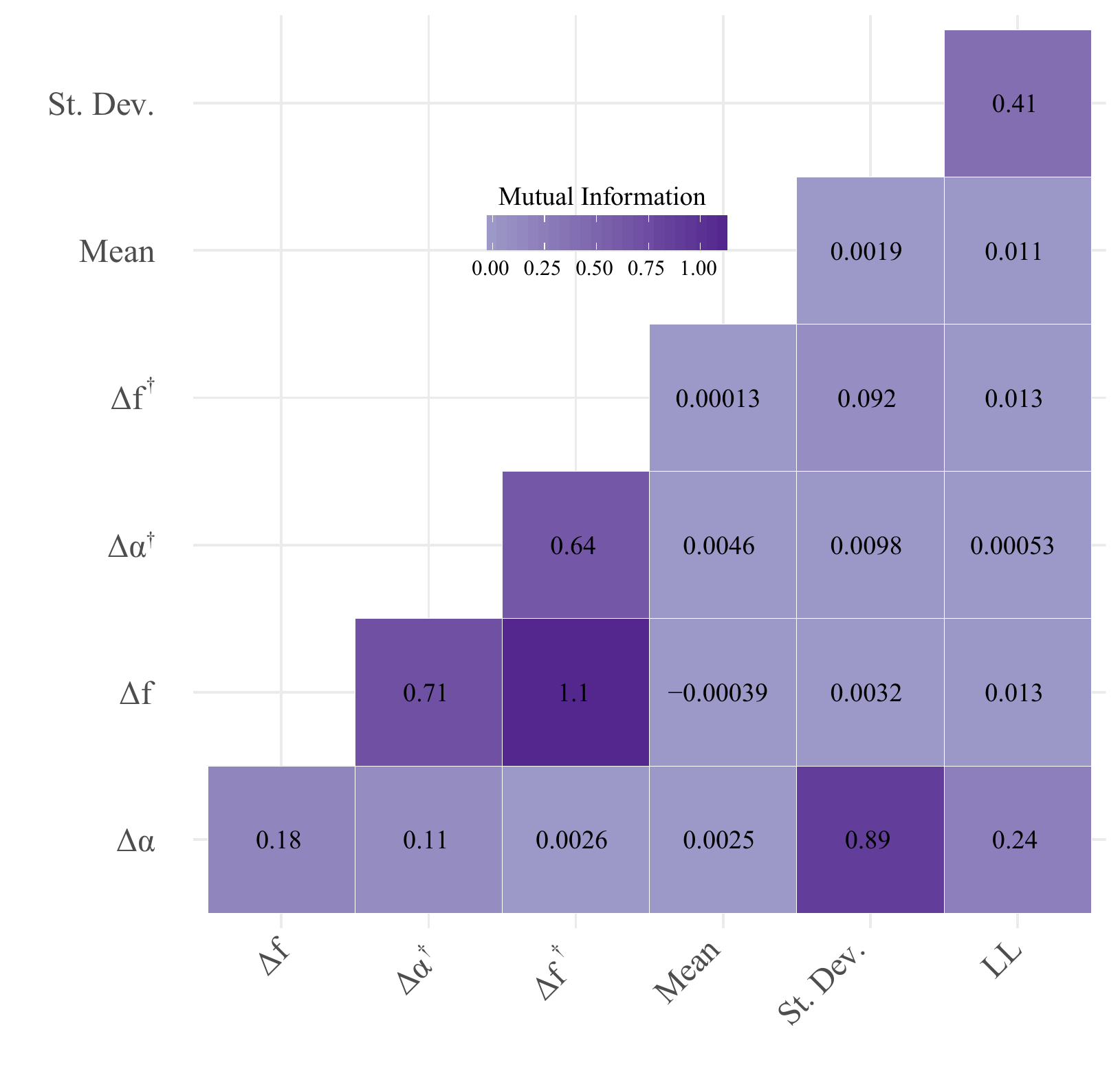}
\caption{{\bf Mutual Information of different measures.}  The MI measure shows higher values for the pair $\Delta\alpha$/St. Dev. than to $\Delta\alpha^{\dagger}$/St. Dev., similar to the result in Fig. 5.
}
\end{figure}
\clearpage

\section*{S7 Fig}
\begin{figure}[h]
\includegraphics[width=\textwidth]{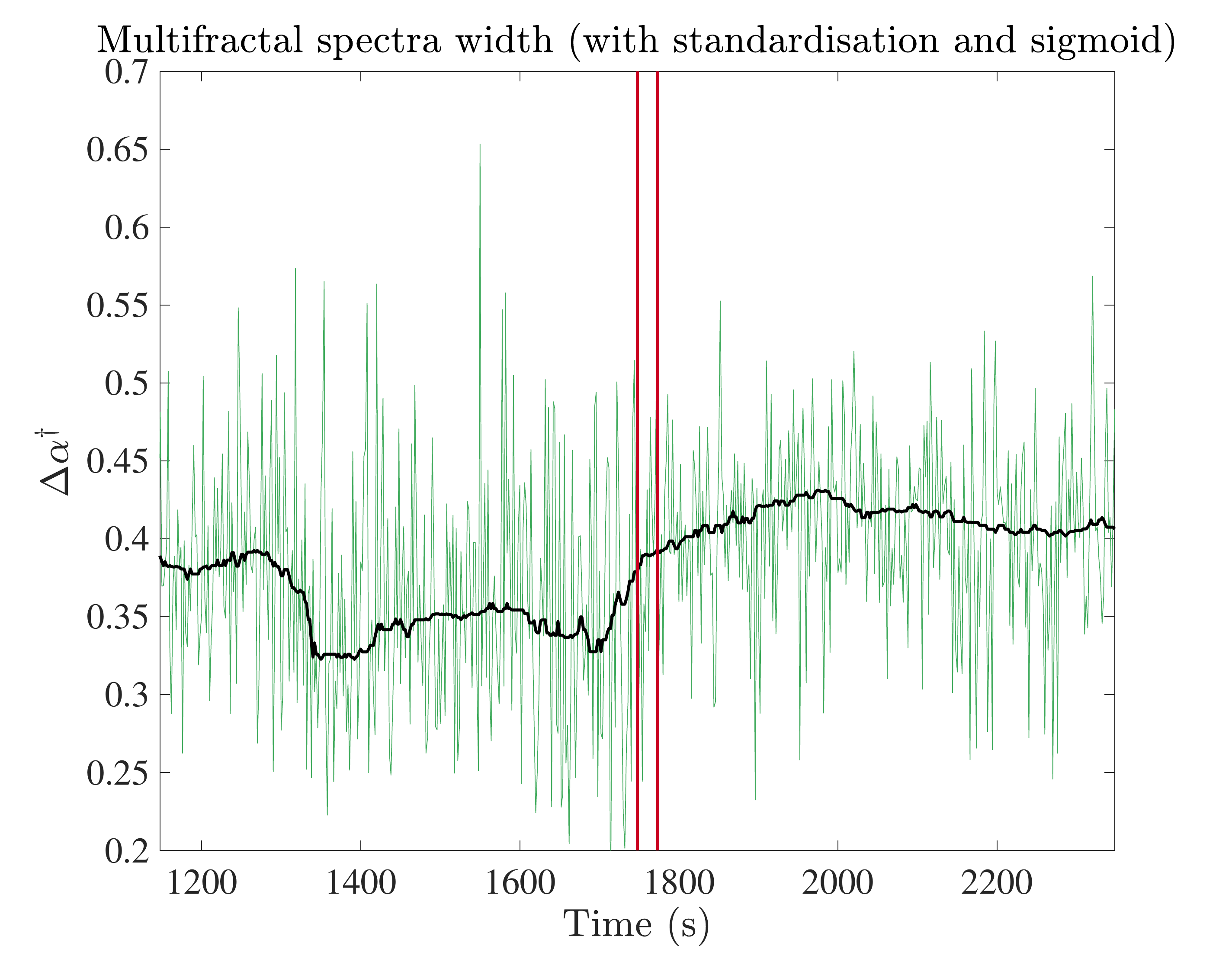}
\caption{{\bf Multifractal spectrum width ($\Delta\alpha^{\dagger}$) for a single icEEG channel in the patient NHNN1 (channel 1) around a seizure.} Black line: moving average.
}
\end{figure}

\bibliography{ref}

\end{document}